\documentclass[reprint,amsmath,amssymb,aps]{revtex4-2}

\newcommand{\RomanNumeralCaps}[1]
    {\MakeUppercase{\romannumeral #1}}

\usepackage{graphicx}
\usepackage{hyperref}
\usepackage{siunitx}
\usepackage{dcolumn}
\usepackage{bm}
\usepackage{color}
\DeclareUnicodeCharacter{2212}{-}

\begin{document}

\preprint{APS/123-QED}

\title{Strong Correlations and Superconductivity in the Supermoiré Lattice}

\author{Zekang Zhou$^{1}$}
\author{Cheng Shen$^{1}$}
\author{Kry\v{s}tof Kol\'{a}\v{r}$^{2}$}
\author{Kenji Watanabe$^{3}$}
\author{Takashi Taniguchi$^{4}$}
\author{Cyprian Lewandowski$^{5,6}$}
\author{Mitali Banerjee$^{1,7, \dagger}$}

\affiliation{$^{1}$Institute of Physics, Ecole Polytechnique Fédérale de Lausanne, CH-1015 Lausanne, Switzerland}

\affiliation{$^{2}$\mbox{Dahlem Center for Complex Quantum Systems and Fachbereich Physik, Freie Universit\"at Berlin, 14195 Berlin, Germany}}

\affiliation{$^{3}$Research Center for Functional Materials, National Institute for Materials Science, 1-1 Namiki, Tsukuba 305-0044, Japan}

\affiliation{$^{4}$International Center for Materials Nanoarchitectonics, National Institute for Materials Science, 1-1 Namiki, Tsukuba 305-0044, Japan}

 \affiliation{$^{5}$National High Magnetic Field Laboratory, Tallahassee, FL 32310, USA}
 \affiliation{$^{6}$Department of Physics, Florida State University, Tallahassee, FL 32306, USA}

 \affiliation{$^{7}$Center for Quantum Science and Engineering, Ecole Polytechnique Fédérale de Lausanne, CH-1015 Lausanne, Switzerland}
 
\affiliation{$^{\dagger}$ Email: mitali.banerjee@epfl.ch}

\begin{abstract}
The supermoiré lattice, arising from the interference of multiple moiré patterns, dramatically reshapes the electronic band structure by introducing new minibands and modifying band dispersion. Concurrently, strong electronic interactions within moiré flat bands lead to the emergence of various correlated states. However, the impact of the supermoiré lattice on the flat band systems with strong interactions remains largely unexplored. Here, we report the existence of the supermoiré lattice in the mirror-symmetry-broken twisted trilayer graphene, elucidating its role in generating mini-flat bands and mini-Dirac bands. Furthermore, we demonstrate interaction-induced symmetry-broken phases in the supermoiré mini-flat bands alongside the cascade of superconductor-insulator transitions enabled by the supermoiré lattice. Our work shows that robust superconductivity can exist in the mirror-symmetry-broken TTG and underscores the significance of the supermoiré lattice as an additional degree of freedom for tuning the electronic properties in twisted multilayer systems, sheds light on the correlated quantum phases such as superconductivity in the original moiré flat bands, and highlights the potential of using the supermoiré lattice to design and simulate novel quantum phases.
 
\end{abstract}

\maketitle

\begin{figure*}
  \centering
  \includegraphics[width= 0.9\textwidth]{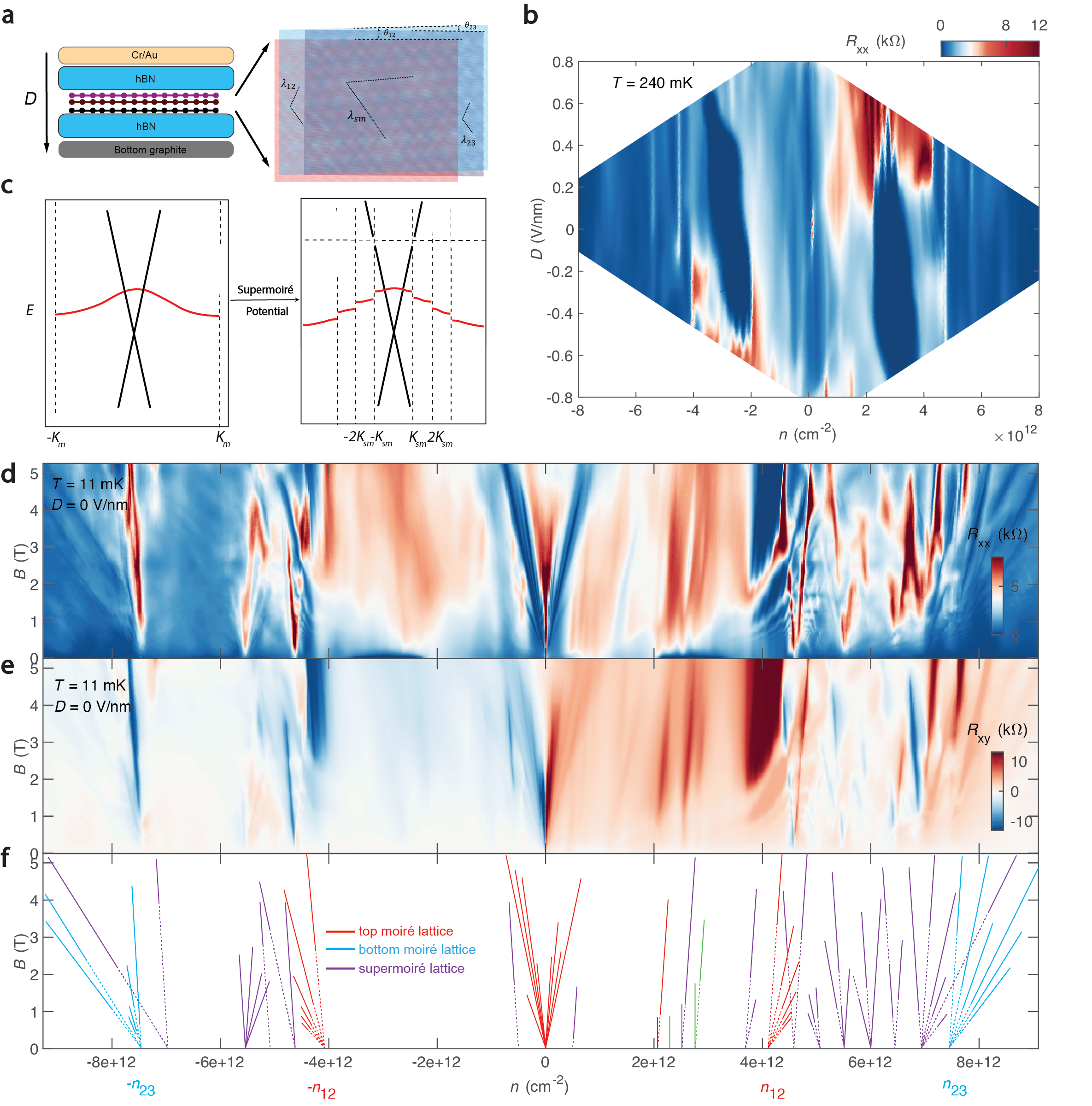}
  \caption{\textbf{Mirror-symmetry-broken twisted trilayer graphene.} \textbf{a}, Schematic of the device with double gates to tune carrier density and displacement field independently, and the right figure illustrates the moiré pattern of the mirror-symmetry-broken TTG, consisting of two small moiré lengths $\lambda_{12}$, $\lambda_{23}$ and a large supermoiré length $\lambda_{sm}$. \textbf{b}, Longitudinal resistance $R_{xx}$ as a function of $n$ and $D$ measured at $T$ = 240~mK. $R_{xx}$ shows zero resistance and many resistance states in a large region in the $n-D$ mapping. \textbf{c}, Schematic illustrating the impact of the supermoiré lattice on the band structure. The left side shows the sketch of the band structure of TTG; the flat bands coexist with a dispersive Dirac band, and the Fermi velocity of the Dirac band is far larger than that of the flat band. The right side shows the band structure after band folding by the supermoiré lattice. The original moiré flat band is diced into mini-flat bands in the supermoiré Brillouin zone. Also, the Dirac band is folded to form the satellite Dirac points. \textbf{d,e}, Landau fan diagram of $R_{xx}$ and $R_{xy}$ at $D$ = 0~V/nm at $T$ = 11~mK. \textbf{f}, States in \textbf{d,e}. Blue lines denote the Landau fan emanating from the full filling of the bottom moiré lattice $\pm n_{23}$; red lines denote the Landau fan stemming from CNP, half-filling correlated state, and band insulators of the top moiré pattern; purple lines represent Landau levels from supermoiré lattice-induced states.}
\label{fig1}
\end{figure*}

\section{Introduction}
The family of 2D moiré materials has recently emerged as a near-perfect platform for discovering and systematically exploring a wide range of quantum states of matter. Beginning with the observation of fractal physics of Hofstadter butterfly~\cite{Dean2013May, Hunt2013Jun, Ponomarenko2013May, Yankowitz2012May, Spanton2018Apr}, the study of moiré systems now has extended to the realization and simulation of intriguing correlated quantum states involving superconductivity and topological (fractional) quantum anomalous/spin Hall effect, etc~\cite{Cao2018Apr, Cao2018Apr1, Yankowitz2019Mar, Wu2021Apr, Saito2021Apr, Das2021Jun, Park2021Apr, Nuckolls2020Dec, Choi2021Jan, Pierce2021Nov, Xie2021Dec, Yu2022Jul, Tang2020Mar,Zhao2023Apr,Tao2024Mar, Xiong2023May,Ciorciaro2023Nov, Li2021Dec,Park2023Oct, Xu2023Sep, Kang2024Apr, Zhao2024Feb, Lu2024Feb, Xia2024Oct}. The tunability of the 2D systems allowed us to consistently push the frontiers of studying moiré physics by introducing new experimental methods of control. Here, we focus on one such technique: the realization of a supermoiré pattern. 

In the presence of multiple moiré lattices, the moiré systems can exhibit a secondary interference process between the original moiré lattices, which gives rise to the formation of the supermoiré lattice. This interference, in turn, gives rise to an effective, even larger-scale supermoiré potential that can modify the original moiré lattice symmetry and underlying band structure~\cite{Wang2019Apr, Finney2019Nov, Wang2019Dec, Yang2020Dec, Kuiri2021Mar, Sun2021Dec, Hu2023Jul, Jat2024Mar}. Despite the relatively simple experimental requirements for the appearance of supermoiré patterns, its impact on the correlated physics present in the moiré systems remains largely unexplored~\cite{Zhu2020Sep, Nakatsuji2023Oct, Popov2023Oct, Yang2024Sep}. Motivated by this observation, we focus on one such natural platform that gives rise to the supermoiré lattice - the alternating twisted trilayer graphene (TTG).

The general form of the alternating angle TTG can be obtained by stacking and twisting three layers of graphene, where the relative angles $\theta_{1,2}$ and $\theta_{2,3}$ between the layers $1$ and $2$ and layers $2$ and $3$ satisfy $\theta_{1,2}\theta_{2,3} < 0$~\cite{Zhu2020Sep, Nakatsuji2023Oct}. Such a system features two moiré patterns with moir\'e wavevectors, $\mathbf g_{12} = \theta_{12} \times \mathbf g$ and $\mathbf g_{23} = \theta_{23} \times \mathbf g$, and wavelengths $\lambda_{12} = a/\theta_{12}$ and $\lambda_{23} = a/\theta_{23}$ (Here the graphene reciprocal lattice vector is $\mathbf g= 4\pi/(\sqrt{3}a) \mathbf {\hat y}$ and $a=$ 0.246 nm represents the lattice constant of graphene). A special and most frequently studied configuration of TTG is the symmetric configuration $\theta_{12}=-\theta_{23}$, where the single-particle electronic spectrum has exactly one twisted bilayer graphene-like flat band which is decoupled from a dispersive Dirac cone~\cite{Khalaf2019Aug, Zhu2020Sep}.  When $\theta_{12}=-\theta_{23}\approx\sqrt{2}\theta_{\text{magic}}$ (here $\theta_{\text{magic}}$ is the magic-angle of twisted bilayer graphene)~\cite{Khalaf2019Aug, Zhu2020Sep}, the system exhibits overall similar experimental phenomenology to the twisted bilayer graphene featuring interaction-driven insulating phases and superconductivity~\cite{Park2021Feb, Hao2021Mar, Cao2021Jul, Shen2023Mar, Liu2022May, Zhou2024Apr}. Away from the symmetric twist angle condition, if the two moiré lattices are close to commensurate, satisfying $p \theta_{12} \approx q \theta_{23} $, where $p$ and $q$ are integers, their interference results in an approximate supermoiré lattice with an effective moiré wavevector given by $|\mathbf g_{sm}| \approx |\mathbf g_{12}|/q \approx |\mathbf g_{23}|/p$. For $p$ and $q$ consecutive integers, the new supermoir\'e lengthscale can be written as $\lambda_{sm}=a/\theta_{13}$ where $a$ is the lattice constant and $\theta_{13}$ is the twist angle between the outer layers.
Note that beyond this leading order supermoir\'e lattice, there is a secondary modulation arising from $|p  \mathbf g_{12} - q \mathbf g_{23}| \neq 0$. This modulation occurs at very long wavelengths and its impact is difficult to detect experimentally.

The presence of the supermoiré lattice has been observed in twisted trilayer systems through techniques such as conductive atomic force microscopy, scanning tunneling microscopy, scanning transmission electron microscopy, and thermodynamic measurements~\cite{Li2022Aug, Turkel2022Apr, Craig2024Mar, VanWinkle2024Mar, Park2024Feb, Hao2024Sep, Xie2024Apr, Hesp2024Dec}. 

This work explores the intricate interplay between the supermoiré lattice, the original moiré flat bands with strong electronic interactions, and dispersive Dirac bands. We confirm the presence of the supermoiré lattice through electronic transport measurements by observing the Brown-Zak oscillation and Hofstadter's butterfly stemming from the formation of supermoiré mini-flat bands and mini-Dirac bands. Moreover, we observe isospin symmetry-broken phases within the supermoiré mini-flat bands and robust superconductivity strongly modulated by the supermoiré potential. Our work highlights the importance of the supermoiré lattice in reshaping the electronic band structure of twisted multilayer systems and offers fresh insights into the correlated physics of moiré systems.

\begin{figure*}
  \centering
  \includegraphics[width= 0.95\textwidth]{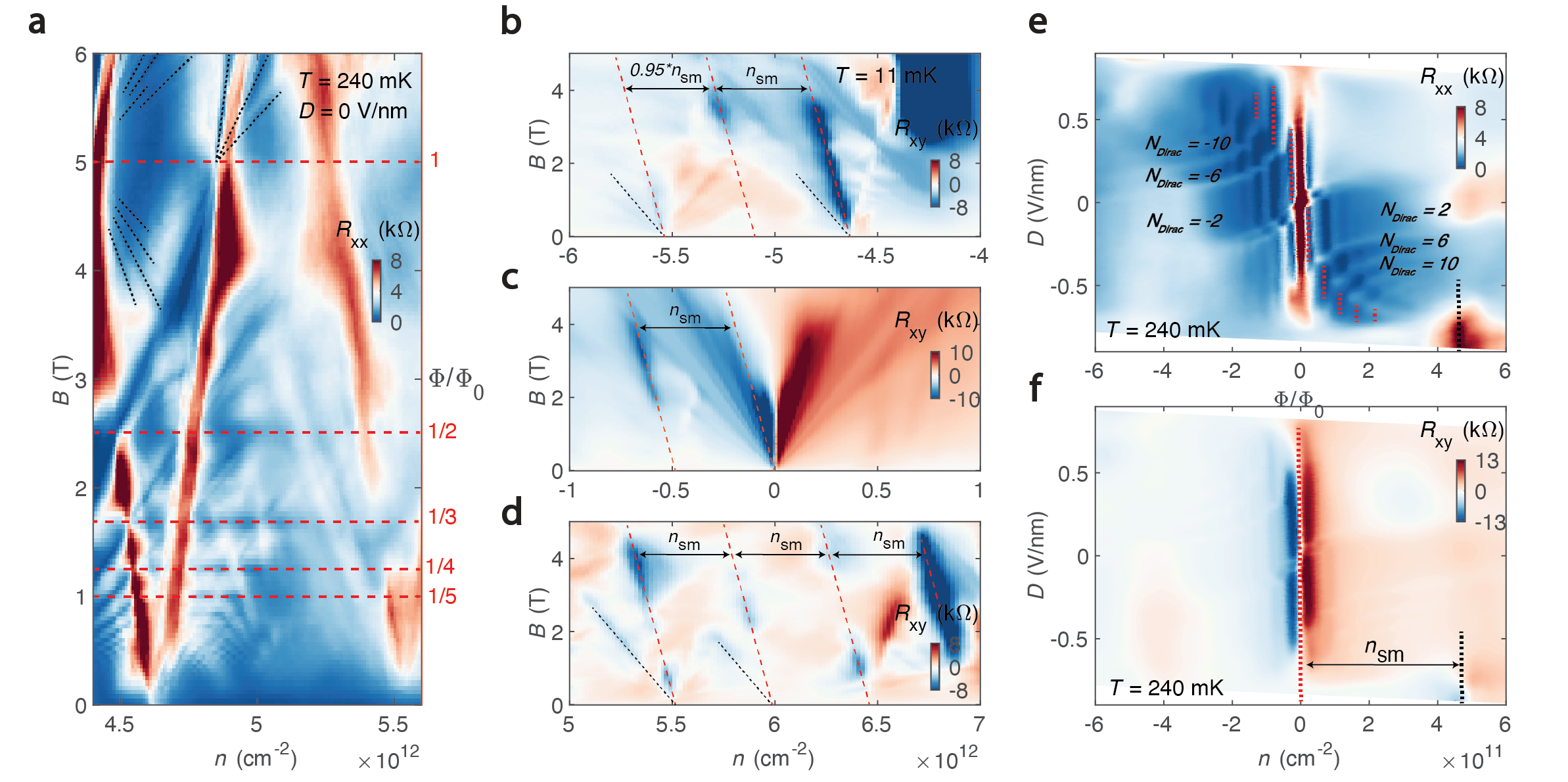} \caption{\textbf{Supermoiré lattice-induced Brown-Zak oscillation, Hofstadter's butterfly and mini-flat bands and mini-Dirac bands.} \textbf{a}, Landau fan diagram of $R_{xx}$ at $D=0V/nm$ shows Hofstadter’s butterfly and Brown-Zak oscillations induced by the supermoiré lattice. The red dashed line marks $1,1/2,1/3,1/4,1/5$ quantum flux of the supermoiré lattice. The black dashed line shows Landau levels originating from one quantum flux. \textbf{b,c,d}, Zoom-in of Landau fan diagram on the hole side between $-n_{23}$ and $-n_{12}$ (\textbf{b}); near CNP (\textbf{c}); on the electron side between $n_{12}$ and $n_{23}$ (\textbf{d}). The red dashed line marks the pronounced states, and the distance between states is close to $N\cdot n_{sm}$; this signifies the existence of supermoiré mini-bands. \textbf{e,f}, $R_{xx}$ and $R_{xy}$ as a function of $n$ and $D$ close to CNP when $B=0.5$~T. The red dashed line tracks the CNP of flat bands, and the black dashed line marks the state emerging from Dirac band folding.}
  \label{fig2}
\end{figure*} 

\section{Mirror-symmetry-broken twisted trilayer graphene}

Fig.~\ref{fig1}a illustrates the double gate geometry of the device, where the carrier density $n=(C_{tg}V_{tg}+C_{bg}V_{bg})/e$ and displacement field $D=(C_{tg}V_{tg}-C_{bg}V_{bg})/2\varepsilon_0$ can be tuned independently ($\varepsilon_0$ is vacuum permittivity, $V_{tg}$/$V_{bg}$) is the applied top/bottom gate voltages and $C_{tg}$/$C_{bg}$) is the capacitance of the top/bottom gate) (See Methods for the device fabrication process). In a typical TTG device in which $\theta_{12}=-\theta_{23}$, mirror symmetry is expected~\cite{Khalaf2019Aug, Zhu2020Sep}, leading to a symmetric $n-D$ phase diagram with respect to the $D=0$ V/nm line~\cite{Park2021Feb, Hao2021Mar, Cao2021Jul, Shen2023Mar, Liu2022May, Zhou2024Apr} (See also Fig.~\ref{Extended1}a,b, $\nu-D$ mapping for a mirror-symmetric TTG device with twist angle $1.35^\circ$). However, our measurements in Fig.~\ref{fig1}b show asymmetrical results of $R_{xx}$ between $D$ and $-D$, indicating the absence of mirror symmetry. Taken together with the absence of TTG and hBN alignment, this suggests that $\theta_{12}\neq -\theta_{23}$ and mirror symmetry broken in the device. This conclusion is further supported by $n - D$ map in Fig.~\ref{fig1}b, which shows a large number of resistive peaks and zero-resistance (superconducting) domes spanning across a broad charge density region both on the electron and hole sides. This is in stark contrast to the mirror-symmetric TTG, where the onset of correlated states is well defined by integer moiré fillings (See Fig.~\ref{Extended1}), 

To thoroughly characterize the resistive states and confirm the twist angles present in the device, we conduct an out-of-plane magnetic field dependence measurement. Fig.~\ref{fig1} d and Fig.~\ref{fig1} e present the Landau fan diagrams in $ R_{xx} $ and $ R_{xy} $ at $ D=0$ V/nm, with a summary in Fig.~\ref{fig1} f. We categorize the observed Landau fans into three sets: states due to the bottom moiré lattice, states hosted by the top moiré lattice, and those arising from the supermoiré lattice. From the full filling carrier density of the single moiré lattice, twist angles can be extracted by using the expression $n_f=8\theta^2/\sqrt{3}a^2$, yielding: $\theta_{12} = 1.328^\circ$ and $\theta_{23} = -1.785^\circ$ (Here $n_f$ is the full filling carrier density of the single moiré lattice. Fig.~\ref{sub3} provides more details on how to define the full filling carrier density of two moiré lattices). The extracted unequal twist angles confirm the lack of mirror symmetry and the presence of two moiré patterns within the system hinted at by Fig.~\ref{fig1}b. Notably, several Landau fans are incommensurate with the full filling set by the top or bottom moiré lattice, as marked by the purple lines shown in Fig.~\ref{fig1}f. We interpret them as Landau-level states stemming from the emergent supermoiré mini-bands, see Fig.~\ref{fig1}c.

Before proceeding with further analysis, we want to pause and place our work in the larger context of previously reported results in the literature on the alternating angle TTG away from the symmetric twist angle condition. In Ref~\cite{uri2023superconductivity}, the two twist angles are $1.4^\circ$ and $-1.9^\circ$ with the superconductivity (and other electronic transport measurements) interpreted as emerging from the moiré quasicrystal formation rather than the formation of a supermoiré lattice. In Ref~\cite{Xie2024Apr}, on the other hand, the authors report on the presence of a supermoiré lattice with the twist angle difference of $0.43^\circ$. The system is studied using compressibility measurements and presents multiple incompressible states attributed to the presence of the supermoiré lattice. Our measurements thus complement these existing experiments reporting on the transport manifestation of the isospin symmetry breaking and observation of superconductivity in the supermoiré regime.

\section{Supermoiré lattice Hofstadter's butterfly and supermoiré mini-bands}

The ratio of the two moiré lattice lengths in our system is close to $3/4$, indicating that the moiré lattices are commensurate. The interference between them can therefore give rise to a well-defined supermoiré lattice. Indeed, the presence of the supermoiré lattice in our system can be confirmed with the low-field Brown-Zak oscillations~\cite{Brown1964Feb, Zak1964Jun}, which exhibit conductance oscillations arising from the interplay between the magnetic flux and the supermoiré potential. Fig.~\ref{fig2}a is the zoom-in of the Landau fan diagram of $R_{xx}$ near full filling density $n_{12}$. $R_{xx}$ dips (conductance peaks) at constant magnetic field are observed at $B$ =2.5~T, 1.7~T, 1.25~T, 1~T as marked by red dashed lines, which corresponds to $1/2, 1/3, 1/4, 1/5$ quantum flux of the unit cell with a spatial period $\lambda_{sm-exp}=30.6$~nm (See also Fig.~\ref{Extended3} and Fig.~\ref{Extended11} for more data). This length scale is in agreement with the expected supermoiré length of $\lambda_{sm-cal}=a/(\theta_{13})=a/(\theta_{12}+\theta_{23})=30.8$~nm. As shown in the supplementary materials, we have ruled out the existence of a moiré lattice between hBN and graphene. In addition to observing Brown-Zak oscillation, the Landau-level states stemming from $5$~T toward both the high magnetic field side and the low magnetic field side (black dashed lines in Fig.~\ref{fig2}a) confirm the identification of $B$ = 5T as corresponding to one quantum flux within the supermoiré unit cell.

Similar to the physical picture of band folding by the original moiré lattice, the supermoiré lattice will further fold and tailor the moiré bands into mini-bands, as depicted in Fig.~\ref{fig1}c. As a result of the secondary band folding, we expect the appearance of additional Landau fans. To search for this effect, we look more closely at Fig.~\ref{fig2}b,c,d of the Hall resistance $R_{xy}$ in the region between the full filling densities of $-n_{23}$ and $-n_{12}$ on the hole side (b), near CNP (c) and in the region between $n_{12}$ and $n_{23}$ on the electron side (d).

In Fig.~\ref{fig2}b, three sets of Landau fans are visible, as marked by the dashed lines. The charge density spacing between these sets of Landau levels is $0.95\cdot n_{sm}$ and $n_{sm}$ (where $n_{sm}=4/(\sqrt{3}/2 \cdot \lambda_{sm}^2)=4.83\times10^{11}$~cm$^{-2}$ is the full filling carrier density of supermoiré mini-bands); in Fig.~\ref{fig2}c, in addition to the Landau fan originating from CNP, another state marked by the red dashed line appears in a finite magnetic field, and the charge density spacing between this state and CNP also corresponds to $n_{sm}$; in Fig.~\ref{fig2}d, four sets of Landau fans exist in this region, and the charge density spacing between each is $n_{sm}$. Furthermore, as shown in Fig.~\ref{Extended2}, $R_{xx}$ shows repeated oscillations with charge density equal to $n_{sm}$. All these observations indicate that supermoiré potential indeed folds further the original moiré bands into mini-bands. We note that the carrier density between supermoiré lattice–induced states can exhibit small variations from case to case. While the precise origin of this small variation is not fully clear, we include error bars in $n_{sm}$ and define full filling as $n_{sm}(1\pm5\%)$. In the following text, $n_{sm}$ includes this small uncertainty.

In addition to the supermoiré potential's role on the original moiré flat bands, it is interesting to explore the fate of the Dirac cone in this trilayer system. As explained earlier, at the symmetric angle condition, TTG is expected to host a Dirac cone that is decoupled from the flat bands. We can thus use a Dirac cone in our system as a ``perturbation'' to the Dirac cone of the symmetric TTG. Alternatively, we can think of the Dirac cone in our system as the original Dirac cone of the (say) third layer when brought in proximity to the moiré system of layers one and two. In either case, the supermoiré potential is expected to fold the Dirac band, producing satellite Dirac points resembling the graphene/hBN moiré superlattice~\cite{Dean2013May, Hunt2013Jun, Ponomarenko2013May, Spanton2018Apr} and hybridizing the original moiré flat band with the dispersive Dirac cone - see Fig.~\ref{ExtendedTheory}. 

The satellite Dirac points have much higher energy than the flat bands at $D=0$~V/nm since the Dirac band Fermi velocity is close to $10^6$~m/s~\cite{Kim2022Jun}, as illustrated in Fig.~\ref{fig1}c and Fig.~\ref{ExtendedTheory}. We refer to satellite Dirac points as the secondary Dirac points that result from band folding, which may be further gapped by the supermoiré potential. By tuning the displacement field, we can change the energy of the satellite Dirac points compared to the Fermi level, allowing us to identify the satellite Dirac points induced by the supermoiré potential 
- Fig.~\ref{fig2}e,f show $R_{xx}$ and $R_{xy}$ as a function of $n$ and $D$ at $B=0.5$~T near CNP. The results show vertical zero $R_{xx}$ features intersected by multiple `S' shape curves, which implies multi-band electronic transport behavior in the system. We attribute the vertical features to the Landau-level states originating from flat bands. Charge density spacing between each vertical line equals $4eB/h$, which suggests the four-fold degeneracy due to the spin and valley degree of freedom in the flat bands (The degeneracy can be lifted by increasing the magnetic field, as shown in Fig.~\ref{Extended4}). The `S' shape features, however, are the Dirac Landau levels transition lines, between which $N_{Dirac} = \pm10, \pm6, \pm2$ Dirac Landau levels are filled. At a constant total carrier density, varying the displacement field results in a shift (adjustment) of the electron density within the Dirac band, transitioning from $2eB/h$ to $6eB/h$ to $10eB/h$$\ldots$, while the carrier density within the flat bands undergoes corresponding shifts. This process occurs as the displacement field adjusts the relative energy between the Landau-level states of the flat bands and those of the Dirac band. When Hall resistance $R_{xy}$ is quantized at $R_{xy}=h/(N_{Dirac}e^2)$ in the region where $N_{Dirac}$ Dirac Landau levels are filled, the contribution of Hall density from flat bands is zero. This fact, in turn, indicates that the Fermi level is at the CNP or in the correlated gaps of the flat bands, allowing us to track the CNP of the flat bands in the $n-D$ sweep (the red dashed line in Fig.~\ref{fig2}e). Remarkably, along the trace of CNP of flat bands, one $R_{xx}$ peak emerges with a sign change of the corresponding $R_{xy}$ when the carrier density is equal to $n_{sm}$ at $D=-0.6$~V/nm as shown by the black dashed lines in Fig.~\ref{fig2}e,f indicating a satellite Dirac point.

\begin{figure*}
  \centering
  \includegraphics[width= 0.9\textwidth]{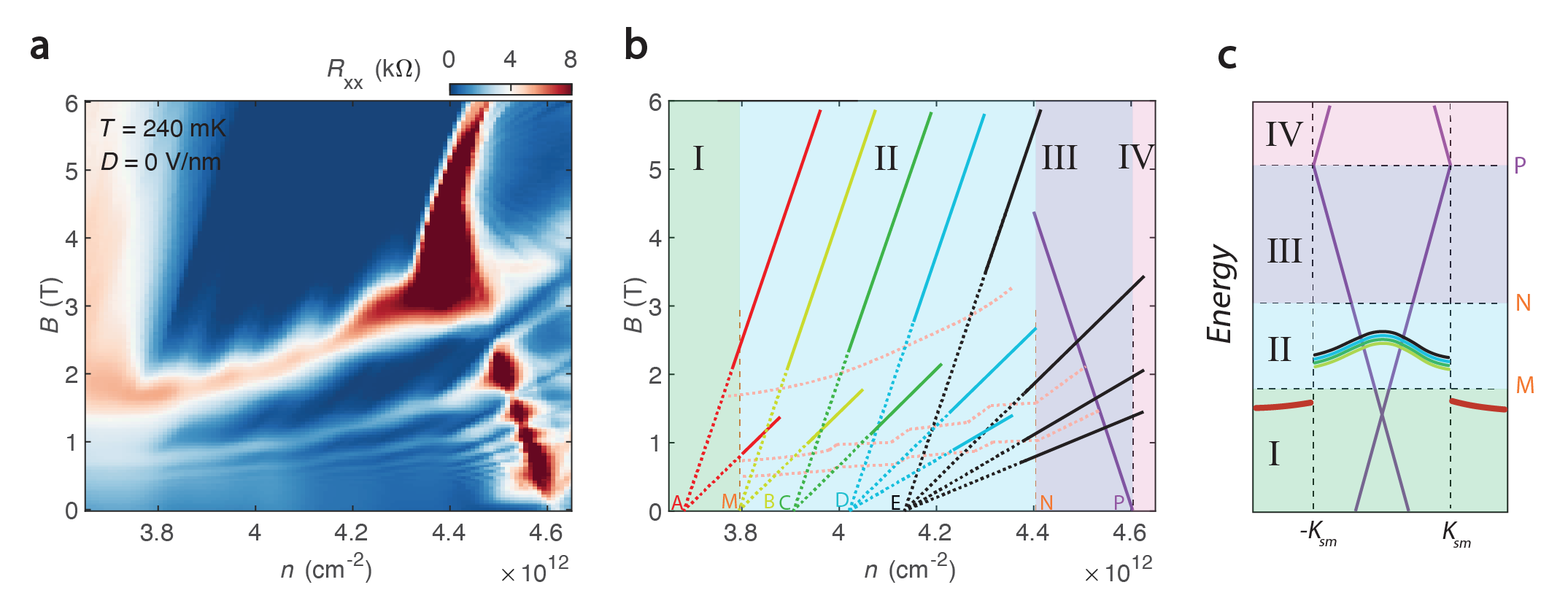}
  \caption{\textbf{Interaction induced isospin symmetry-broken phases in mini-flat bands.} \textbf{a}, Landau fan diagram of $R_{xx}$ at $D=0V/nm$ near the full-filling $n_{12}$ of the top moiré lattice. \textbf{b}, Guiding lines of the gapped states appearing in the $R_{xx}$ map in \textbf{a}. Black lines mark the Landau fan from full filling of the top moiré lattice. And four sets of Landau fans are marked by red, yellow, green, and cyan lines, each with slopes of 2 and 6. The spacing between these sets of Landau fans is kept equal at 0.238$\times n_{sm}$. The purple line shows the satellite Dirac cone. \textbf{c}, sketch of band structure in \textbf{a,b}.
  }
\label{fig3}
\end{figure*}

\begin{figure*}
  \centering
  \includegraphics[width= 1\textwidth]{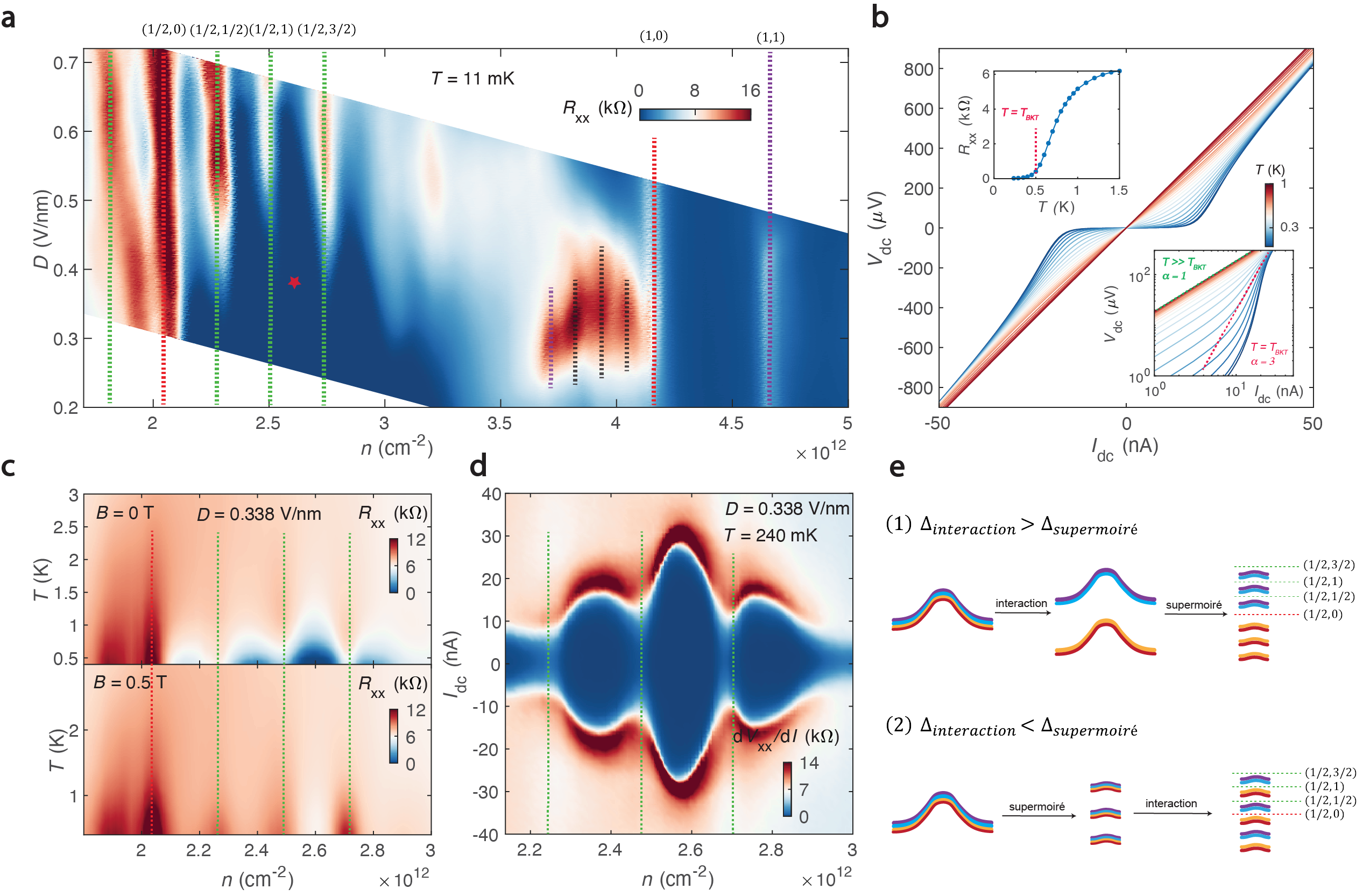}
  \caption{\textbf{Cascade of superconductor-insulator transitions induced by the supermoiré lattice.} \textbf{a}, zoom in of $R_{xx}$ from Fig.~\ref{fig1}b as a function of $n$ and $D$ on the electron side. Red dashed lines mark the half-filled and full-filled states of the top moiré flat bands; green dashed lines mark the half-filled or full-filled states of the supermoiré mini-bands. We also marked these states by using two numbers $(x,y)$ when the states appear at the carrier density $n=1/2\cdot x\cdot n_{12} + 1/2\cdot y \cdot n_{sm}$. \textbf{b}, $V_{dc}-I_{dc}$ curve as a function of the temperature of the red star point in \textbf{a}. The right bottom insert shows $V_{dc}-I_{dc}$ on a log-log scale, $V_{dc}$ follows a high-power polynomial to a linear dependence on $I_{dc}$ as temperature increases. The green line shows $V_{dc} \propto I_{dc}$ when $T\gg T_{BKT}$; the red line shows $V_{dc} \propto I_{dc}^3$, indicating a BKT transition at this temperature. The top left insert shows $R_{xx}$ as a function of $T$ while the red dashed line is the $T_{BKT}$ extracted above. \textbf{c}, $R_{xx}$ versus $n$ and $T$ when $B=0$~T (top panel) and $B=0.5$~T (bottom panel) at $D=0.338$~V/nm. The superconducting dome is divided into several small domes in high displacement fields. High resistance states emerge between superconducting regions. Temperature dependence of $R_{xx}$ of these states at $B=0.5$~T indicates the insulating behavior. The distance between these insulating states corresponds to the half-full filling of the supermoiré lattice. \textbf{d}, $dV_{xx}/dI$ as a function of $I_{dc}$ and $n$ at $D=0.338$~V/nm. Three individual superconducting domes are visible. \textbf{e}, Two scenarios for the half-filled supermoiré insulator, with each color representing a different isospin flavor. In the first scenario, the interaction within the moiré band is stronger than the supermoiré effect. The strong interactions first break the fourfold degeneracy, after which the supermoiré potential divides the moiré band into supermoiré mini-flat bands. In the second scenario, the supermoiré potential dominates over the interaction. The supermoiré potential first divides the moiré band into a set of supermoiré bands, each with four-fold degeneracy. Then the strong interaction within the supermoiré mini-bands lifts the fourfold degeneracy.
  }
\label{fig4}
\end{figure*}

\section{Interaction-induced isospin symmetry-broken phases in supermoiré mini-flat bands}

Having confirmed the presence of secondary supermoiré band folding, we now discuss the role of interaction effects in these mini-bands. The original moiré flat bands exhibit correlated electronic phases due to the dominant Coulomb interactions over kinetic energy. Supermoiré bands further quench the kinetic energy, but since the same supermoiré length scale factor lowers the characteristic Coulomb energy scale, the ratio between Coulomb interactions and the bandwidth of the supermoiré mini-flat bands remains similar to that of the original moiré flat bands. Therefore, mini-flat bands can also host correlated phases. To demonstrate the presence of symmetry-broken states, we will now consider the $R_{xx}$ versus $n$ and $B$ near full filling of the top moiré lattice $n_{12}$ on the electron side, see Fig.~\ref{fig3}a.

The phase diagram in Fig.~\ref{fig3}a can be understood by considering the role of interactions in shaping the interplay between supermoiré lattice, moiré flat bands, and Dirac bands. Fig.~\ref{fig3}b illustrates all the states appearing in Fig.~\ref{fig3}a, and Fig.~\ref{fig3}c shows the idealized sketch of the band structure in this region for clarity of the discussion. The $n$ vs $B$ sweep of $R_{xx}$ in Fig.~\ref{fig3}a can be divided into four regions based on the position of Fermi level, which we mark in Fig.~\ref{fig3}b,c with \RomanNumeralCaps{1}, \RomanNumeralCaps{2}, \RomanNumeralCaps{3} and \RomanNumeralCaps{4}. 

In region \RomanNumeralCaps{3} at point N, the Fermi level lies at the top of the top moiré flat bands. However, the electron density at point N does not equal the full-filling carrier density of the top moiré lattice $n_{12}$ because some electrons reside in the Dirac band. As the electron density increases, the Fermi level enters the moiré band gap of flat bands, and all electrons enter the Dirac band. This regime manifests as clear Dirac Landau levels with a sequence $N_{Dirac}=2, 6, 10,\ldots$ visible as the black lines in Fig.~\ref{fig3}b. Extending these Dirac Landau levels to zero magnetic fields defines the point E, which gives the full-filling electron density of the top moiré lattice $n_{12}=n_E$. When the electron density continuously increases, the Fermi level will eventually lie at the satellite Dirac point (point P). A resistive $R_{xx}$ peak appears at point P, and Landau levels also emanate from point P. The Dirac band electron density at point P is $n=n_{P}-n_{E}$, which exactly equals $n_{sm}$, confirming the nature of the point P as the satellite Dirac
point. 

In region \RomanNumeralCaps{2}, both the Dirac band and flat bands are partially occupied, with the majority of electrons residing in the flat bands owing to their enhanced density of states. Transitions between different Dirac Landau levels are visible, as shown by the dashed pink lines in Fig.~\ref{fig3}b (Fig.~\ref{Extended5}a,b shows the corresponding $R_{xy}$). At point M, a gap opens in the moiré flat bands due to the supermoiré potential, and an isolated supermoiré mini-flat band with four-fold degeneracy is formed. The supermoiré gapped state at point M exhibits a slope of $N_{Dirac}$ in the fan diagram, represented by the red lines. Extending the Landau levels to zero magnetic field defines the point A corresponding to the actual electron density $n_A$ filled into flat bands at point M (We point out that we ignore the small contribution from the increase of Dirac band carrier density when the Fermi level moves across the gap). The electron density difference in moiré flat bands between point M (where Fermi level lies in the supermoiré gap) and point N (where Fermi level lies at the top of flat bands) is $n_{E}-n_{A}$, is also equal to $n_{sm}$, which suggests the states marked by the red lines are indeed the single-particle band gap state of supermoiré mini-band. 

In the $R_{xx}$ diagram of Fig.~\ref{fig3}a,b, three groups of Landau fans are marked by yellow, green, and cyan lines that are not accounted for by single-particle physics. We find that these Landau fans' zero-field charge density intercepts are equally spaced with a charge spacing corresponding to a quarter-full filling $1/4\cdot n_{sm}$ of the supermoiré mini-flat band. We interpret these states as representing isospin symmetry-broken states arising from interactions within the supermoiré mini-band. In analogy to the earlier analysis (see also Ref. \cite{Shen2023Mar}), the slope of Landau levels from these mini-band states at quarter filling is sensitive to the filling of the Landau levels from the Dirac band. The pink dashed lines also display subtle jumps across the symmetry-broken states in the mini-flat bands, confirming the gap opening at these quarter fillings~\cite{Shen2023Mar, Jiang2025}. These results point to the crucial role the supermoiré potential can play in promoting symmetry-breaking. As expected, in the absence of a supermoiré and additional band folding near the full-filling of moiré flat bands, isospin symmetry breaking does not occur. However, when the bands are further modified by the presence of an additional translational symmetry breaking (here, the supermoiré lattice), it can enable a new set of isospin-symmetry-broken phases.

We briefly comment on the apparent robustness of the decoupled Dirac band to hybridization with the mini-bands. Carrying out a standard continuum model analysis 
\cite{moraGuerciChernMosaicIdeal2024,fuDevakulMagicangleHelicalTrilayer2023,moraMaoSupermoireLowenergyEffective2023,tarnopolskyPopovMagicAngleButterfly2023,guineaFooExtendedMagicPhase2024} to obtain local band structures, see Fig. \ref{ExtendedTheory}, suggests that away from the symmetric twist angle condition, a strong hybridization between the mini-bands and the Dirac cone should exist. As such, one would not expect a well-defined Dirac cone. The presence of a Dirac cone is in agreement with the prior work of Ref.~\cite{uri2023superconductivity}, for which it has been proposed to arise due to the Coulomb interaction~\cite{Yang2024Sep}. Here we propose that this robustness of the Dirac cone could be explained by an unequal tunneling strength between the layers $1$ and $2$, and the layers $2$ and $3$ (see Fig.~\ref{ExtendedTheory}). This reasoning is in line with recent theoretical works~\cite{Nakatsuji2023Oct} that suggest that alternating TTG tends to relax to symmetric TTG conditions, thus accounting for the robustness of the Dirac cone. Lastly, we also caution in passing that the Ref. \cite{Yang2024Sep} argues that in the case of twist angles present in our device rather than supermoiré unit cell formation, one should expect the presence of a quasicrystal formation. Such a crystal would exhibit variation of regions of local commensurability on a length scale of ($>100$ nm). Each locally commensurable region would however, in principle still give rise also to a length scale similar to what we define as the supermoiré length (simply due to the commensurate configuration) and thus potentially in agreement with the Brown-Zak oscillations of Fig.\ref{fig2}a. The length scale corresponding to variation of the commensurability however, is too large for us to see experimentally, and thus we cannot comment on that aspect of the Ref. \cite{Yang2024Sep} predictions.

\section{Cascade of superconductor-insulator transitions}

We now proceed to the most exciting impact of the supermoiré on the electronic properties of the TTG moiré graphene. Robust superconductivity has been observed in multiple alternating angle twisted moiré systems in the symmetric twist angle setup; however, observation of superconductivity in TTG away from this twist angle regime remains rare. Previously, it was observed in a moiré quasicrystal TTG device, and we also report it here in the supermoiré TTG device.

The mirror-symmetry-broken twisted trilayer graphene exhibits superconductivity on both the electron and hole sides, as shown in Fig.~\ref{fig4}a. Fig.~\ref{fig4}b shows $V_{dc}$ versus $I_{dc}$ as a function of the temperature of the red star point in Fig.~\ref{fig4}a. The bottom right insert shows the $V_{dc}-I_{dc}$ curve on a log-log scale, revealing that $V_{dc}$ transitions from a high-power polynomial dependence to a linear dependence on $I_{dc}$ as the temperature increases. The red dashed line shows $V_{dc} \propto I_{dc}^3$, signifying a two-dimensional superconducting Berezinskii-Kosterlitz-Thouless (BKT) transition at temperature $T=T_{BKT}=0.5$~K. The left top insert shows $R_{xx}$ as a function of temperature (the red dashed line is the $T_{BKT}$ extracted above), also demonstrating a typical superconductor behavior. 

Most strikingly, Fig.~\ref{fig4}a shows that resistive states extend into the superconducting dome at high displacement fields. Using the previous determination of the full filling carrier density of top moiré lattice $n_{12}$ and the supermoiré lattice $n_{sm}$, we can characterize these resistive states as follows: half-filled correlated states of the top moiré bands ($1/2\cdot n_{12}$) shown by red dash lines; interaction-induced insulators stemming from half filled or fully filled supermoiré mini-bands ($1/2\cdot n_{12}+1/2\cdot N \cdot n_{sm}$, N is positive number) shown by green dashed lines. In contrast to the carrier density interval between resistive states in the high doping region (Fig.~\ref{Extended2}), the carrier density interval between these resistive states corresponds to half-full ($n_{sm}/2$) filling of supermoiré mini-bands. Superconductivity present in the supermoiré mini-bands and top moiré flat band is strongly modulated by these states as shown by the bias current dependence (Fig.~\ref{fig4}c) and temperature dependence (Fig.~\ref{fig4}d), both measured at $D=0.338$~V/nm. 

The presence of intertwining insulators and superconductivity in the region $1/2\cdot n_{12}+1/2\cdot N\cdot n_{sm}$ is a particularly intriguing aspect of our work. We attribute this interplay of the two correlated phenomena to the complex competition between the effective supermoiré potential and interactions in the original moiré flat bands. This competition is illustrated in Fig.~\ref{fig4}e, where we highlight two distinct situations depending on the relative strengths of the supermoiré-induced gap and the interaction-induced gap. 
\begin{enumerate}
\item If the interaction-induced gap is much stronger than the supermoiré effect, then at half-full filling of $n_{12}$, the four-fold spin-valley degeneracy of the 
moiré flat bands is lifted first. Supermoiré mini flat bands, formed by Brillouin zone folding of the two-fold degenerate moiré band, then host the same two-fold isospin degeneracy. In this case, the emergent insulators at $1/2\cdot n_{12}+1/2\cdot N\cdot n_{sm}$ are effectively single particle insulators of the supermoiré flat bands. Each superconducting dome spans an entire mini-flat band and has the same isospin order as others. 

\item If the interaction-induced energy scale is much smaller than the supermoiré gap, then the supermoiré potential gives rise to the moiré flat bands folding first, with each mini-flat band having four-fold isospin degeneracy. This degeneracy is consequently lifted by interactions. In this scenario, the mini-bands exhibit a cascade effect, similar to that in MATBG, but only at half-filling. Specifically, within one supermoiré mini-band, the first half-filling maintains fourfold degeneracy, while the second half-filling reduces to twofold degeneracy except approaching the full filling of the supermoiré band. This implies that superconductivity in this case does not necessarily require isospin symmetry breaking.

\end{enumerate}

It's important to note that the above discussion assumes an identical tendency for symmetry breaking across all the mini-bands. Realistically, we anticipate the different mini-bands to have quantitatively different electronic dispersions, and hence, we argue different propensities for symmetry breaking. This naive argument may account for the observed appearance of correlated insulators only near the half-full filling density region. 

\section*{Discussion and Conclusion}

Our findings highlight the potential of the supermoiré lattice in shaping the band structure and properties of correlated states in moiré systems. The intriguing coexistence of superconductivity with multiple insulating phases suggests that further exploration of the TTG away from the symmetric twist angle condition is needed. We propose that a systematic characterization of the nature of the correlated phases in the supermoiré platform could potentially limit possible pairing mechanisms in the moiré graphene platform. Moreover, the supermoiré lattice is not limited to twisted trilayer graphene but is a universal feature in moiré heterostructures. This opens up a new realm of experimental possibilities, as future multilayer devices focusing on other quantum states, such as the fractional Chern insulating states, could leverage this new degree of experimental freedom to realize yet another set of unexpected electronic quantum phases.

\newpage
\clearpage

\bibliography{supermoire}

\section*{METHODS}

\subsection{Sample fabrication and Measurements}
The stacks are made by using the dry-transfer technique. Graphene, graphite, and hBN are exfoliated on the {SiO$_2$}/Si chips, and the thickness and quality of the materials are checked under an optical microscope. The long monolayer graphene is cut into three pieces before being picked up in order to reduce strain during the stacking process. A homemade poly(bisphenol A carbonate)/polydimethylsiloxane (PC/PDMS) stamp is used to pick up all the flakes one by one. The stack is finally dropped down on the aligned marker chips at 200 $^\circ$C. 5nm/20nm Cr/Au is evaporated on the top of the stack to serve as the top gate and mask of the etch process. 5nm/60nm Cr/Au is evaporated to connect the graphene to form one-dimensional contacts.

Electronic transport measurements are performed in two fridges. One is an Oxford He-3 fridge with a base temperature of $ T=240$mK, and another is a Bluefors dilution refrigerator with a base temperature of $ T=11$mK. Resistance measurements are conducted using a standard lock-in technique employing a 1-10 nA AC current excitation at a frequency of $17$Hz. Two Yokogawa GS200 are used to apply top and bottom gate voltages to tune the carrier density and displacement field. Voltage signals are taken before being amplified 100 times.

\subsection{Extraction of two twist angles}
The twist angle is linked to full moiré carrier density by the relationship $N_f=8\theta^2/\sqrt{3}a^2$, where $a= 0.246~\text{nm}$ is the lattice constant of graphene. In order to get the full filling carrier density, Hall resistance is measured as shown in Fig.\ref{sub2}a. We used antisymmetric $R_{xy} = (R_{xy}(B)+R_{xy}(-B))/2$ to reduce the effect of $R_{xx}$ on $R_{xy}$, Fig.\ref{sub2}b show $R_{xy}$ as a function of $B$ at different $V_{tg}$ and the Hall density at different $V_{tg}$ can be extracted through the relation $R_{xy}=B/(en)$.  The carrier density in the device is determined by $n=(C_{tg}V_{tg}+C_{bg}V_{bg})/e$ and the displacement field is $D=(C_{tg}V_{tg}-C_{bg}V_{bg})/2\varepsilon_0$. In Fig.\ref{sub2}a, we fixed $D=0$, so $V_{bg}=C_{tg}/C_{bg}*V_{tg}$ and carrier density can be expressed as $n=2C_{tg}V_{tg}/e$. Fig.\ref{sub2}c shows Hall density as a function of $V_{tg}$, and the linear fit will give $2C_{tg}$. Two gate voltages $V_{tg(full)}$ and $V_{bg(full)}$ corresponding to the full filling of the flat band in the $R_{xx}(V_{tg}, V_{bg})$ mapping can be obtained. Full carrier density is then expressed as $N_{f}=(C_{tg}V_{tg(full)}+C_{bg}V_{bg(full)}){/e}$. 

\subsection{Single-particle continuum model description of the supermoiré bandstructure}
To model mirror-asymmetric alternating trilayer graphene, we consider a commensurate approximation, taking $\theta_{1,2}=p\theta $ and $\theta_{2,3}= -q\theta $, where our system approximately corresponds to $(p,q)=(3,4)$. Following Ref.~\cite{Popov2023Oct}, we write down the local Hamiltonian for this system in the $K$-valley as follows
\begin{equation}
\label{eq:tbgham}
H_{TBG}^{K}=\begin{pmatrix}
v_F \mathbf k\cdot \boldsymbol{\sigma} +\Delta U & w_{1\leftrightarrow 2}T(p\mathbf r)& 0 \\
w_{1\leftrightarrow 2}T^\dagger(p\mathbf r)&v_F \mathbf k \cdot \boldsymbol{\sigma}&w_{2\leftrightarrow 3}T^\dagger(q\mathbf r - \mathbf d) \\
0&w_{2\leftrightarrow 3}T(q\mathbf r - \mathbf d)&v_F \mathbf k \cdot \boldsymbol{\sigma} -\Delta U
\end{pmatrix},
\end{equation}
where $v_F$ is the graphene Dirac velocity,
$w_{1\leftrightarrow 2}$ ($w_{2\leftrightarrow 3}$) is a dimensionless parameter that allows to tune the tunnelings between the first and second (second and third) layers, $\boldsymbol{\sigma}$ is the vector of Pauli-matrices in sublattice space, and where $\Delta U$ is an interlayer potential difference used to model an external displacement field. The tunneling term reads 
\begin{multline}
\label{eq:tbgtunnel}
T(\mathbf r) = (w_{AA}\sigma_0 + w_{AB}\sigma_x )e^{i \mathbf q_1 \cdot \mathbf r} + \\(w_{AA}\sigma_0 + w_{AB}\sigma_x e^{2\pi i /3 \sigma_z} ) e^{i \mathbf q_2 \cdot \mathbf r}+\\
+ (w_{AA}\sigma_0 + w_{AB}\sigma_x e^{4\pi i /3 \sigma_z} )e^{i \mathbf q_3 \cdot \mathbf r},
\end{multline}
where $w_{AA}$ parametrizes the strength of intrasublattice tunneling and $w_{AB}$ the strength of intersubattice tunneling.
Above $\mathbf q_j = \frac{4\pi \theta}{3 a_0}(O_3)^j  \left[0 ,-1\right] $, where $a_0$ is the monolayer graphene lattice constant,
and $O_3$ the matrix of a counterclockwise $120^\circ$ rotation.
The vector $\mathbf d$ parametrizes~\cite{Khalaf2019Aug}
the displacement of the third layer with respect
to the first layer.

To gain an intuitive understanding of the band structures, we fix the strength of the tunneling between the first and second layers to
the TBG values $w_{1\leftrightarrow 2}=1$, but treat $w_{2\leftrightarrow 3}$ as a tuning parameter. We use $v_F = \SI{542.1}{meV\cdot nm}$,  $w_{AB} = \SI{110}{meV}$, $w_{AA} = 0.7 w_{AB}$, $\mathbf d = \mathbf d_{AB}$, where $\mathbf d_{AB} = \theta a_0 (\frac{1}{2\sqrt{3}},-\frac{1}{2})$ is the location of the AB stacking point for twisted bilayer graphene at twist angle $\theta$. We use the twist angle $\theta = 0.45^\circ$, giving $\theta_{1,2}=1.35^\circ$ and $\theta_{2,3}= -1.8^\circ $, close to the values of our device.To model an applied displacement field, we also vary the interlayer potential difference $\Delta U$. We show the resulting band structures in Fig.~\ref{ExtendedTheory} for four values of $w_{2\leftrightarrow 3}$ and four values of $\Delta U$, showing the increasing hybridization of the third layer Dirac cone for increasing $w_{2\leftrightarrow 3}$, as well as the shifting of the Dirac cone with $\Delta U$. The purpose of this modelling is not to quantitatively reproduce the data, but rather to provide an intuitive explanation for how a robust Dirac cone could arise in a symmetry broken TTG.

\begin{acknowledgments}
We thank Ady Stern, Yuval Oreg, Jose L. Lado, and Klaus Ensslin for the fruitful discussions. Z.Z. acknowledges funding from SNSF. K.K. was supported by Deutsche Forschungsgemeinschaft through CRC 183 (project C02). C.L. was supported by start-up funds from Florida State University and the National High Magnetic Field Laboratory. The National High Magnetic Field Laboratory is supported by the National Science Foundation through NSF/DMR-2128556 and the State of Florida. M.B. acknowledges the support of the SNSF Eccellenza grant No. PCEGP2\_194528, and support from the QuantERA II Programme that has received funding from the European Union’s Horizon 2020 research and innovation program under Grant Agreement No 101017733. K.W. and T.T. acknowledge support from the JSPS KAKENHI (Grant Numbers 20H00354 and 23H02052) and World Premier International Research Center Initiative (WPI), MEXT, Japan.

\end{acknowledgments}

\textbf{\begin{center}Author Contributions\end{center}}

M.B. supervised the project. Z.Z. fabricated the devices, performed measurements, and analyzed the data with the help of C.S. KK and CL conducted the theory calculations. K.W. and T.T. provided the hBN crystals. Z.Z. wrote the paper with input from all authors.

\textbf{\begin{center}Competing interests\end{center}}
The authors declare no competing interests.

\textbf{\begin{center}Data availability\end{center}}
The data supporting the findings of this study are available from the corresponding author upon reasonable request.

\newpage
\clearpage
\onecolumngrid

\section*{Extended figures}

\begin{figure*}[h]
\renewcommand{\thefigure}{E1}
  \centering
  \includegraphics[width= 0.93\textwidth]{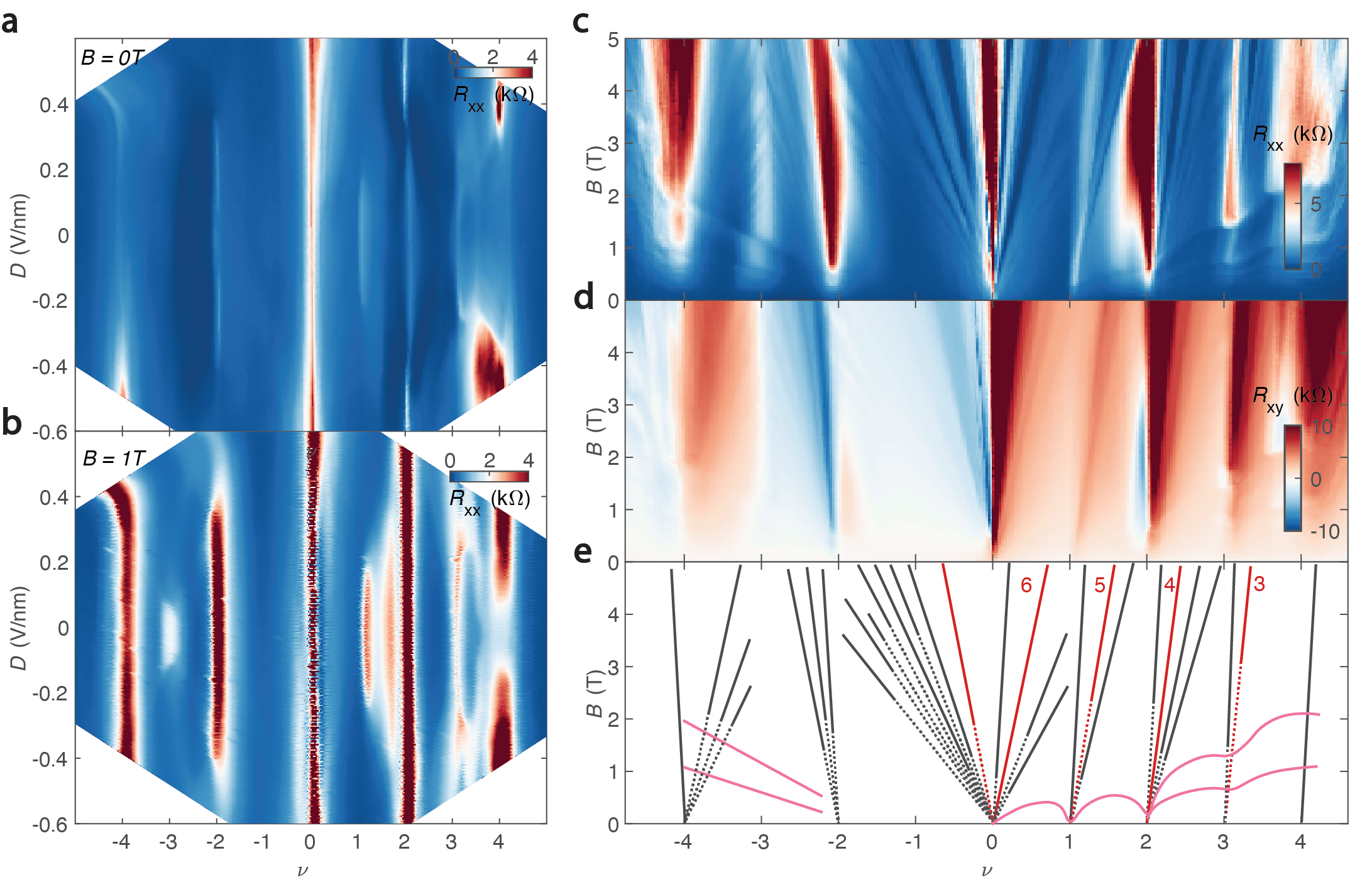}
  \caption{\textbf{Mirror symmetry twisted trilayer graphene device in which $\theta_{12}=-\theta_{23}=1.35^\circ$.} \textbf{a (b)}, $R_{xx}$ as function of $\nu$ and $D$ when $B=0$~T ($B=1$~T) at $T=240$ mK. Superconductivity appears on both the hole-doped side between $\nu=-3$ and $\nu=-2$ and the electron-doped side. Correlated states appear at moiré filling $\nu=1,\pm2,3$. The result is symmetric as a function of the displacement field, which is expected due to the mirror symmetry of the device. \textbf{c (d)}, Landau fan diagram of $R_{xx}$ and $R_{xy}$ when $D=0$~V/nm and the states are extracted and shown in \textbf{e}. Landau levels originating from $\nu=\pm4,0$ and $\nu=1,\pm2,3$ are visible as marked by the black and red lines, while Landau levels from the Dirac band can also be seen as the pink lines. The red lines mark the most robust Landau levels stemming from each correlated state at $\nu$ and have a slope of $C=2+4-\nu$ in the fan diagram in which Dirac Landau levels contribute $C_{Dirac}=2$, and $C_{flat}=4-\nu$ originates from symmetry-broken Chern insulators in flat bands as the same as TBG.}
\label{Extended1}
\end{figure*}

\begin{figure*}[h]
\renewcommand{\thefigure}{E2}
  \centering
  \includegraphics[width= 0.6\textwidth]{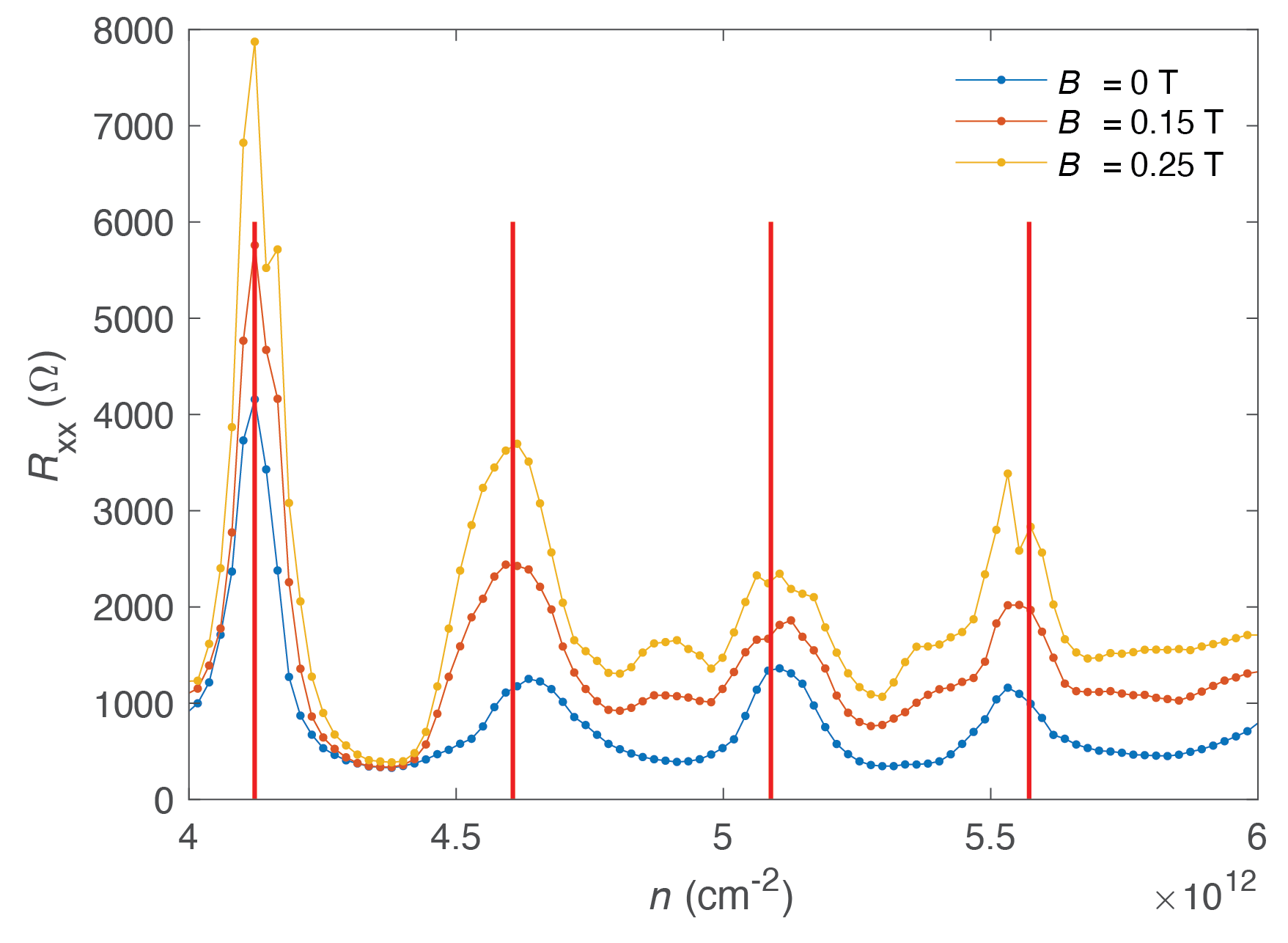}
  \caption{\textbf{Period oscillation of $R_{xx}$ at high doping region.} The longitudinal resistance $R_{xx}$ is plotted as a function of carrier density for a fixed top-gate voltage of $V_{tg} = 11$~V under different magnetic fields. Red lines serve as guides to the eye, indicating the expected positions of resistance peaks. The spacing between successive red lines corresponds to $n_{sm} = 0.468 \times 10^{12}$~cm$^{-2}$. The first resistance peak consistently aligns with the corresponding red guiding line. The second resistance peak shows a slight deviation from the red line; however, at $B = 0.15$~T and $B = 0.25$~T, it aligns well with the guiding lines. The third and fourth resistance peaks exhibit small deviations from the red lines, which we attribute to peak broadening and measurement uncertainties.}
\label{Extended2}
\end{figure*}

\begin{figure*}[h]
\renewcommand{\thefigure}{E3}
  \centering
  \includegraphics[width= 0.95\textwidth]{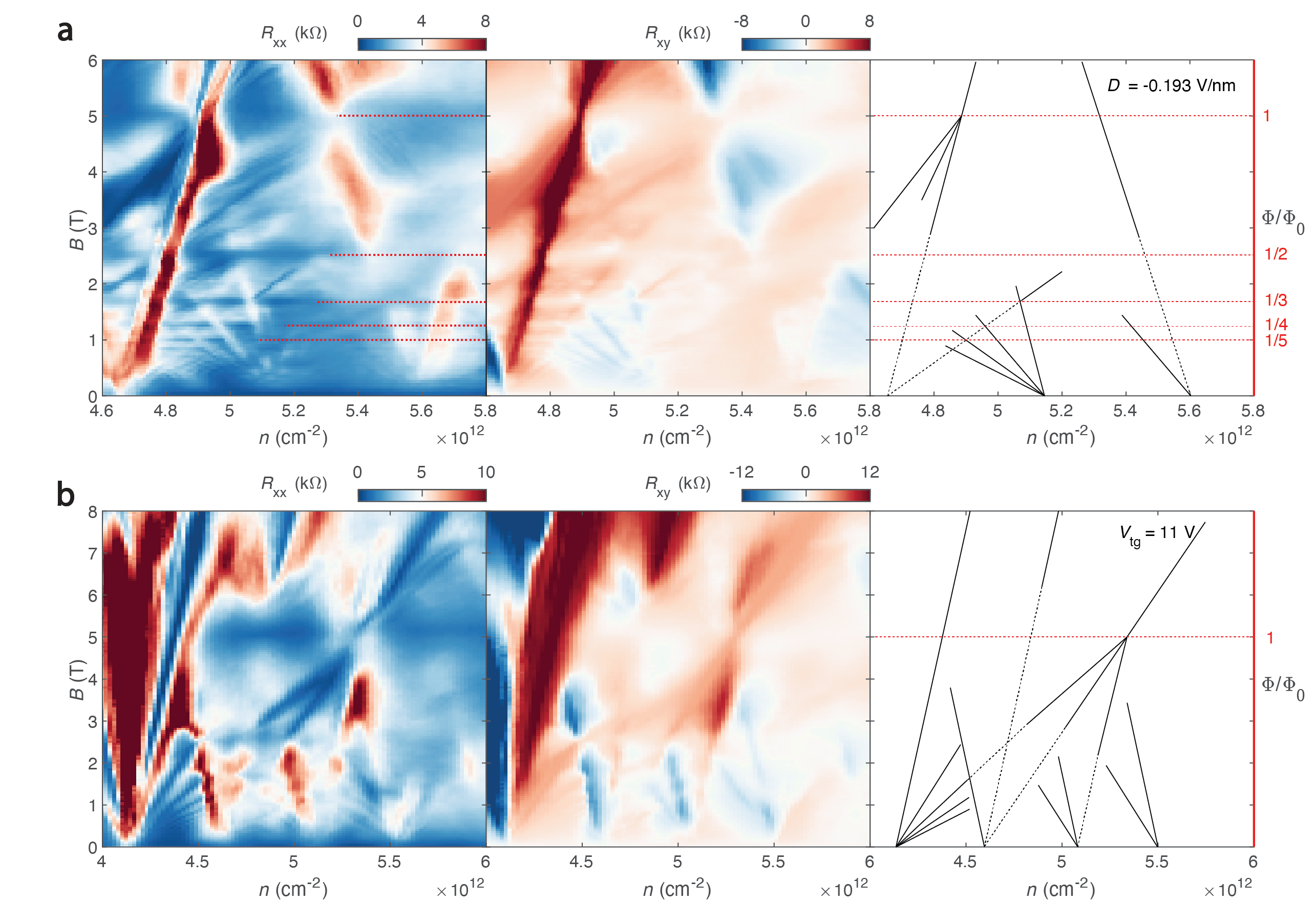}
  \caption{\textbf{Hofstadter's butterfly in other regions of $n-D$ mapping.} \textbf{a}, $R_{xx}$ and $R_{xy}$ as a function of $B$ and $n$ when $D=-0.193$~V/nm. The right figure extracted the states by black lines, and the red lines mark $1,1/2,1/3 \ldots$ quantum flux of the supermoiré lattice. Landau levels stemming from different supermoiré mini-bands are visible. \textbf{b}, the same measurement as \textbf{a} with fixing $V_{tg}=11$~V and varying $V_{bg}$ to change carrier density. Four sets of Landau levels can be seen, and the fan diagram shows Hofstadter's butterfly due to the supermoiré lattice. ($T=11$~mK)}
\label{Extended3}
\end{figure*}

\begin{figure*}[h]
\renewcommand{\thefigure}{E4}
  \centering
  \includegraphics[width= 0.85\textwidth]{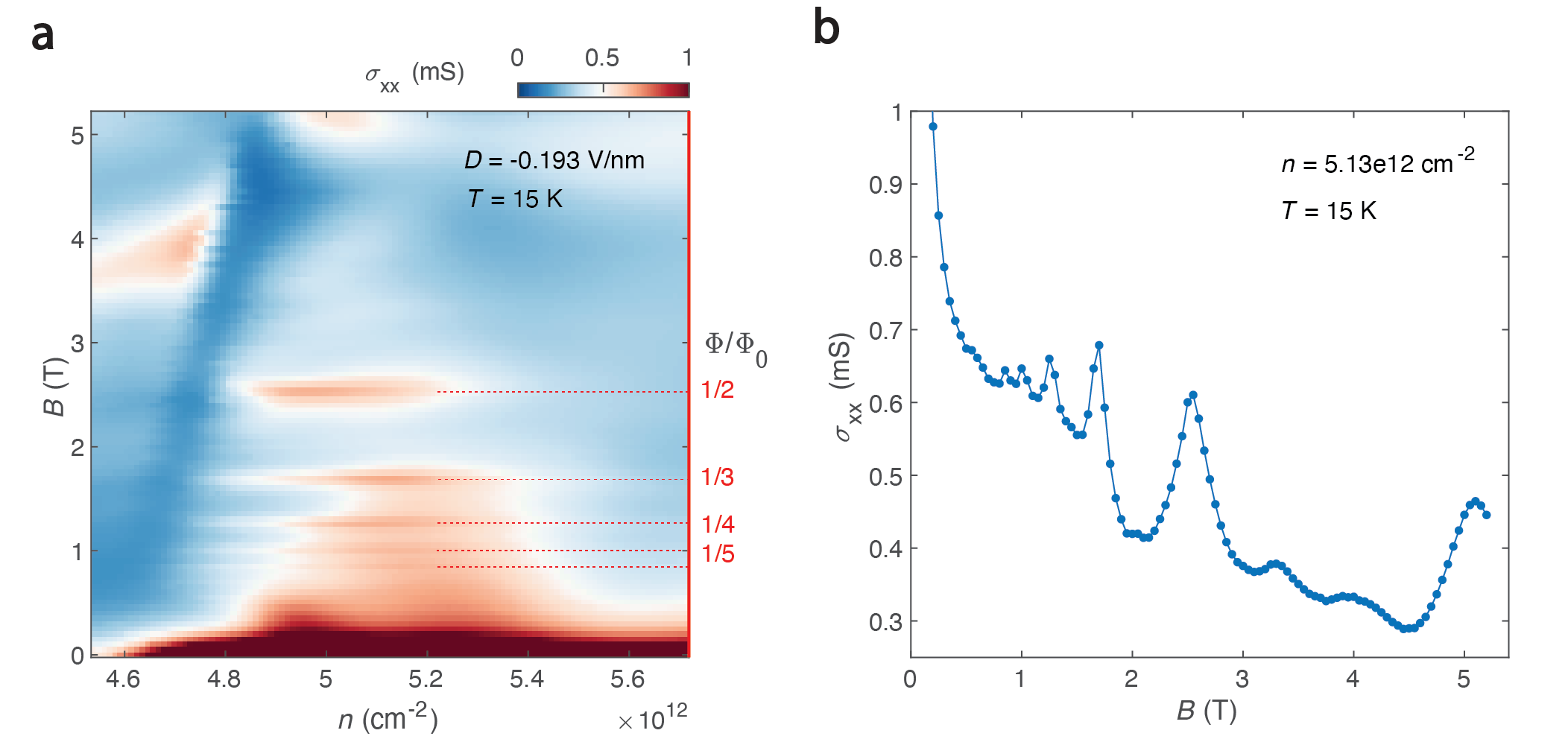}
  \caption{\textbf{Supermoiré Brown-Zak oscillation.} \textbf{a}, longitudinal conductance as a function of $n$ and $B$ at $T=$ 15~K. The conductance shows peaks at the fractional quantum flux of the supermoiré lattice, suggesting the Brown-Zak oscillation. \textbf{b}, linecut of conductance in \textbf{a}.}
\label{Extended11}
\end{figure*}

\begin{figure*}[h]
\renewcommand{\thefigure}{E5}
  \centering
  \includegraphics[width= 0.93\textwidth]{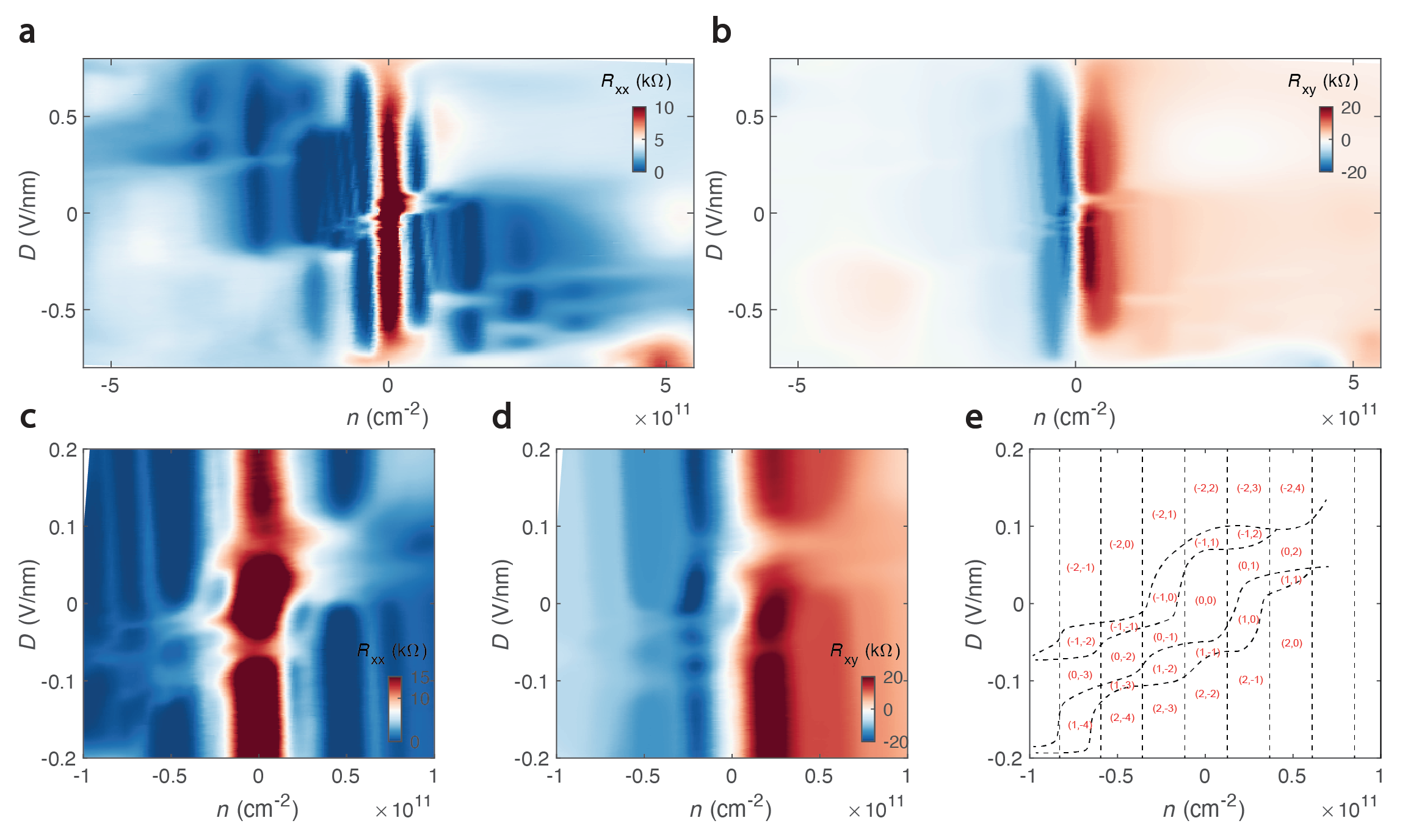}
  \caption{\textbf{Coexisting of Dirac band subsystem and flat band subsystem and co-propagating and counter-propagating edge states near CNP.} \textbf{a}, \textbf{b}, $n-D$ mapping of $R_{xx}$ and $R_{xy}$ when $B=1$~T. Same as in Fig.\ref{fig2}e,f, vertical features are flat band Landau levels, and 'S' shape features are the transition lines of nearby Dirac Landau levels. However, compared with the results measured at $B=0.5$~T, the distance between different vertical features equals $eB/h$, which means the four-fold degeneracy of flat band Landau levels is lifted. We briefly comment on the, the degeneracy of Dirac band Landau levels is also lifted as shown in \textbf{c},\textbf{d}, which are zoom-in of $R_{xx}$ and $R_{xy}$. There are four Dirac Landau level transition lines near CNP, between which $N_{Dirac}$ equals $1,0,-1$. At a fixed $n$, the number of total filled Landau levels $N_{total}=nh/eB$ and the filled flat band Landau levels $N_{flat}=N_{total}-N_{Dirac}$ can be extracted. \textbf{e}, figure shows $(N_{Dirac},N_{flat})$ in the $n-D$ mapping. ($T=240$~mK)}
\label{Extended4}
\end{figure*}

\begin{figure*}[h]
\renewcommand{\thefigure}{E6}
  \centering
  \includegraphics[width= 0.95\textwidth]{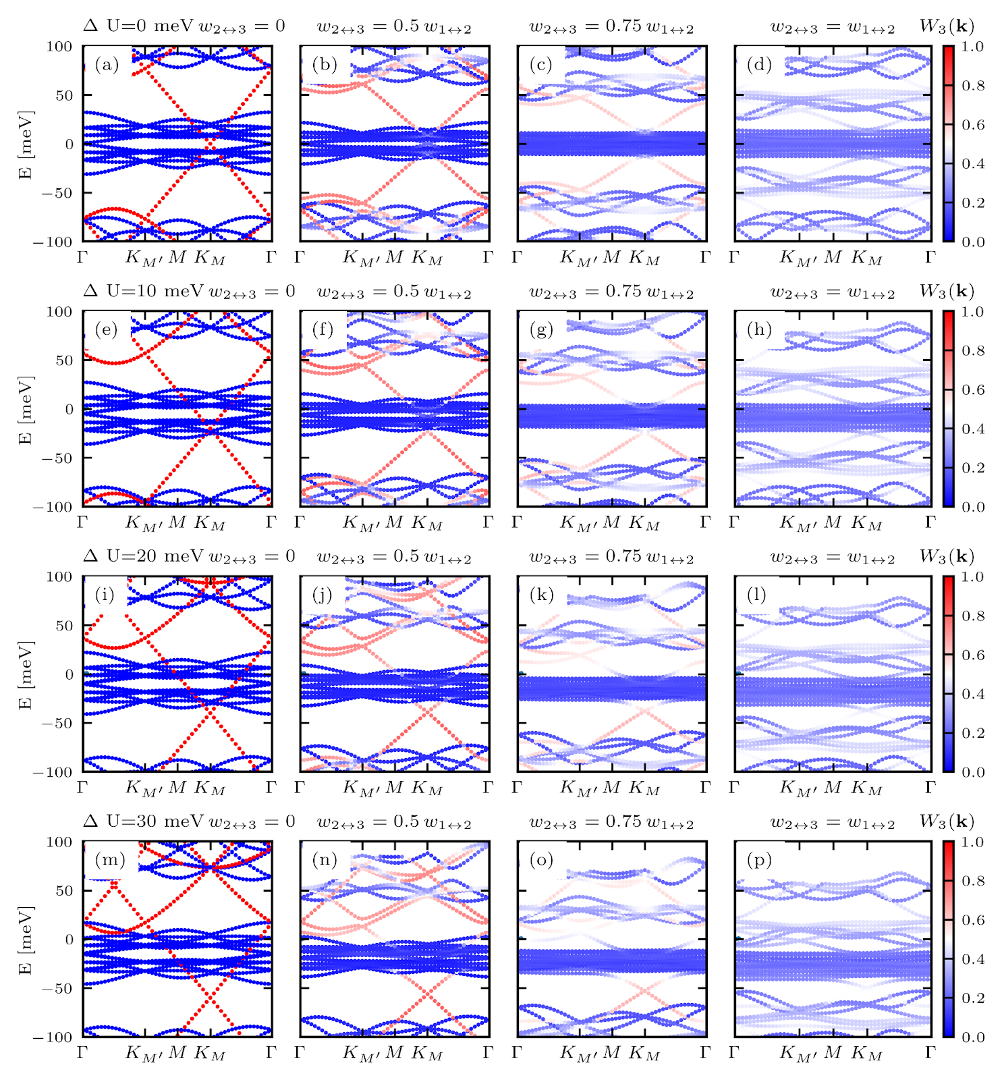}
  \caption{{\bf Hybridization of the Dirac cone with the mini-bands as a function of interlayer tunneling.} We track the single particle electronic structure of the TTG as a function of interlayer potential difference $\Delta U$ and interlayer tunneling between layer 2 and layer 3. The band color indicates the extent of state polarization on the bottom layer (layer 3). In the weak tunneling regime, the Dirac band exists and the relative fermi surface between mini bands and Dirac band can be adjusted by the displacement field. Here we choose zero displacement between the two moiré patterns $d=0$.}

\label{ExtendedTheory}
\end{figure*}

\begin{figure*}[h]
\renewcommand{\thefigure}{E7}
  \centering
  \includegraphics[width= 0.93\textwidth]{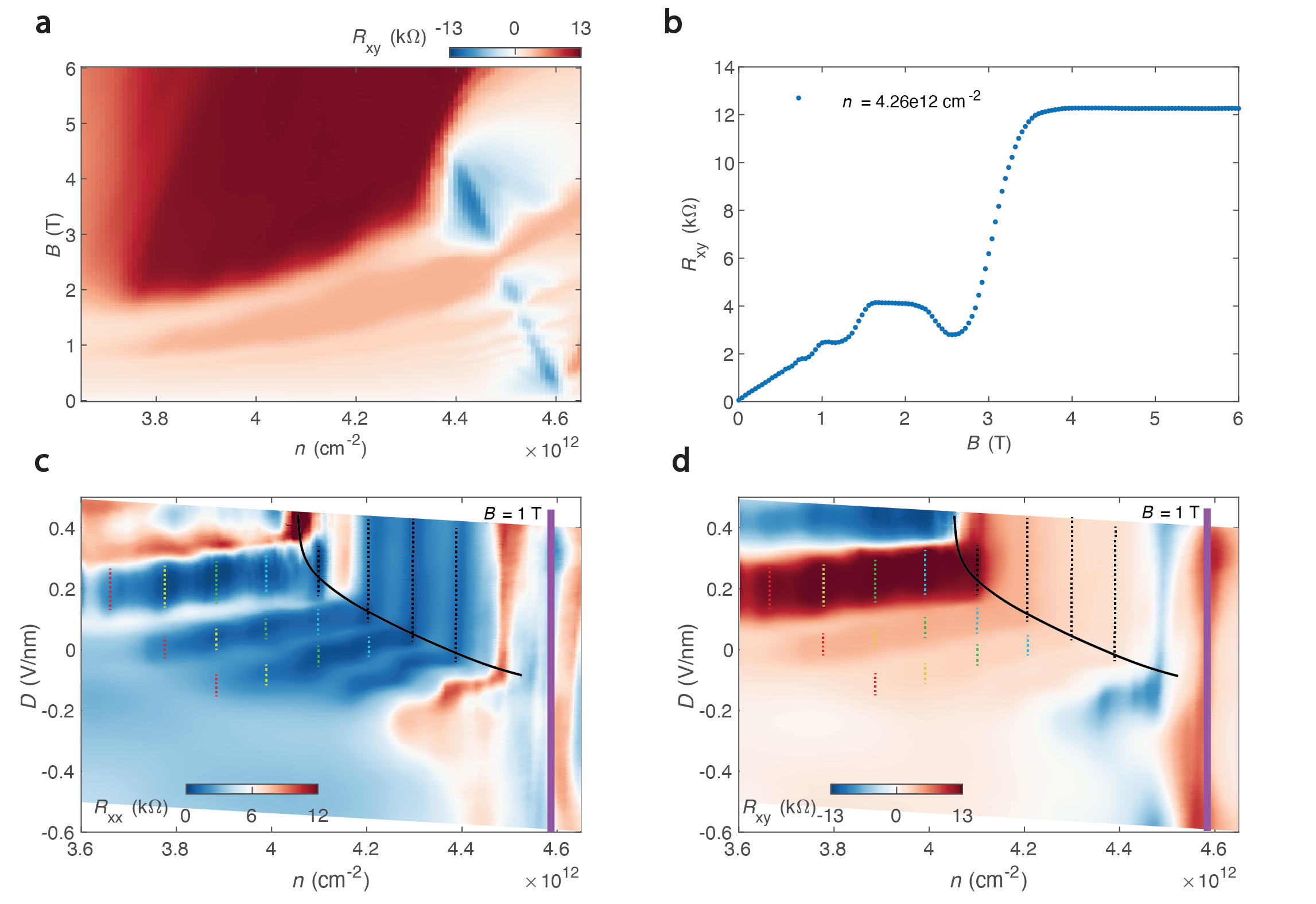}
  \caption{\textbf{Iso-spin symmetry breaking in the supermoiré mini band.} \textbf{a,} The corresponding Landau fan diagram of $R_{xy}$ of the $R_{xx}$ data shown in Fig.~\ref{fig3}a. \textbf{b,} Line cut of $R_{xy}$ at a carrier density of $n = 4.26 \times 10^{12}\ \mathrm{cm}^{-2}$. $R_{xy}$ shows almost quantized resistance values of $h/2e^2$, $h/6e^2$, and $h/10e^2$, attributed to the Dirac Landau levels. When the Fermi level lies within the gap of the flat band, the transport is solely governed by the Dirac band, which transitions into Dirac Landau levels in the presence of a magnetic field. \textbf{c,d,} $n$-$D$ mapping of $R_{xx}$ and $R_{xy}$ at $B = 1$\,T, for electron densities near the full filling of the top moiré lattice. The black solid lines indicate the full filling of the top moiré lattice. It is evident that the position of this full filling shifts as a function of the displacement field. This shift arises from the relative band shift between the Dirac band and the flat band.  When the displacement field becomes sufficiently large (in our case, $D > 0.3\ \mathrm{V/nm}$), the Dirac band hybridizes with the flat band. As a result, $R_{xy}$ no longer exhibits characteristics associated with the Dirac band.  The purple lines mark the strong states appearing at $n_{12} + n_{\mathrm{sm}}$, which occur when the Fermi level lies within the gap of both the Dirac and flat bands. Additionally, the red dashed line highlights a gapped state induced by the supermoiré lattice. Between the red and black lines, three more states—marked by yellow, green, and cyan lines—are observed. These are attributed to isospin symmetry breaking within the supermoiré miniband.
}
\label{Extended5}
\end{figure*}

\begin{figure*}[h]
\renewcommand{\thefigure}{E8}
  \centering
  \includegraphics[width= 0.93\textwidth]{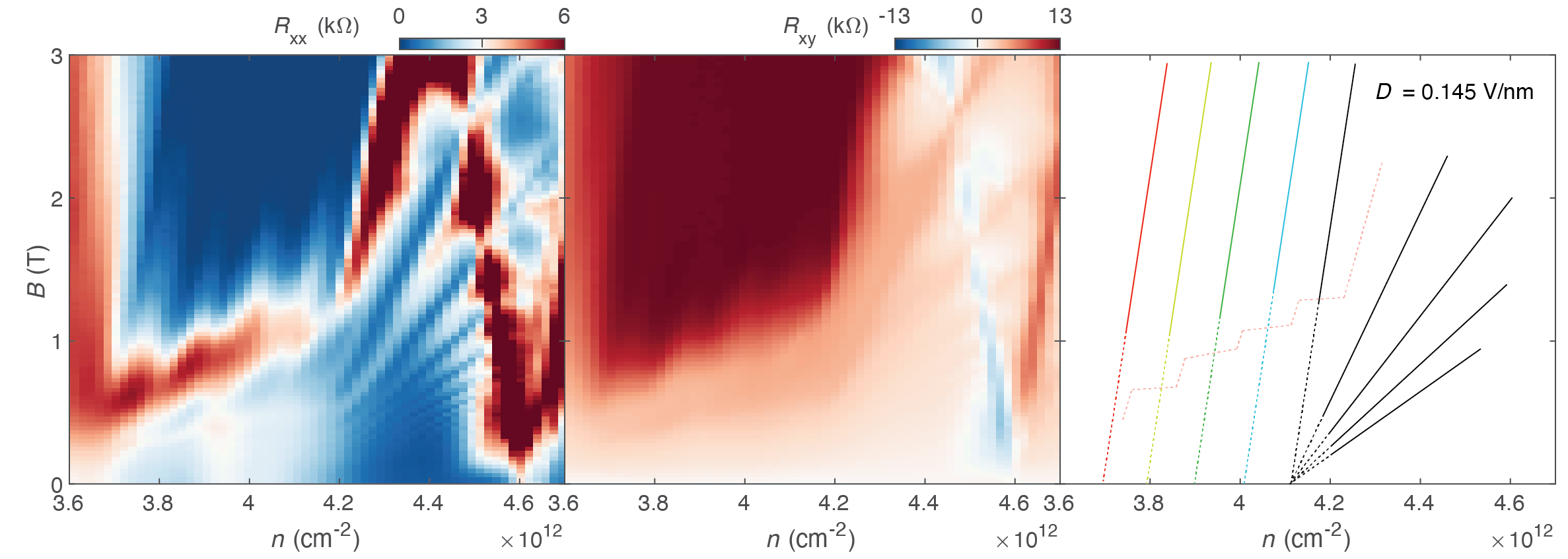}
  \caption{\textbf{Landau fan diagram of $R_{xx}$ and $R_{xy}$ at $D = 0.145~\mathrm{V/nm}$.} All the observed features in the Landau fan are illustrated. The black dashed lines represent Landau levels originating from the full filling of the top moiré lattice ($n_{12}$). The red dashed line marks the gapped state induced by the supermoiré lattice. The pink line indicates the transition between different Dirac Landau level regimes: above this line, two Dirac Landau levels are filled; below it, due to a change in degeneracy, six Dirac Landau levels are filled. Additionally, three states highlighted by yellow, green, and cyan lines are observed, with spacings close to $n_\mathrm{sm}/4$. These are attributed to isospin symmetry breaking within the supermoiré miniband.}
\label{Extended6}
\end{figure*}

\begin{figure*}[h]
\renewcommand{\thefigure}{E9}
  \centering
  \includegraphics[width= 0.75\textwidth]{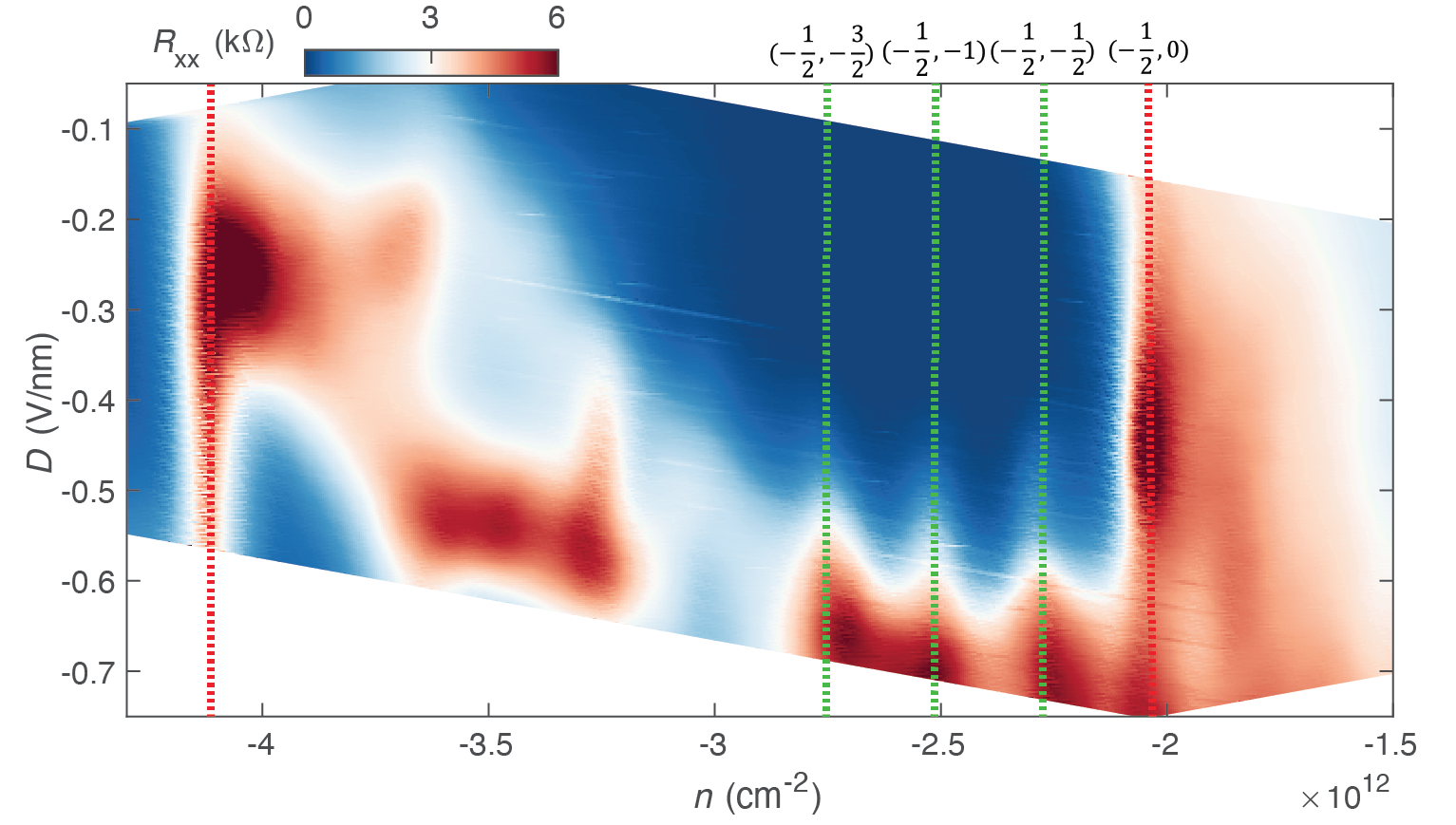}
  \caption{\textbf{$n$-$D$ mapping of $R_{xx}$ on the hole-doped side.} The red dashed lines mark the half filling ($\frac{1}{2}*n_{12}$) and full filling ($n_{12}$) states of the top moiré lattice. The green dashed lines are the guiding lines for the other resistive states appearing in the superconducting regime. And the distance between is $\frac{1}{2}*n_{12}$. Similar to Fig.~\ref{fig4}a, the nature of these states depends on the relative strength between interaction effects and the supermoiré potential.}
\label{Extended9}
\end{figure*}

\begin{figure*}[h]
\renewcommand{\thefigure}{E10}
  \centering
  \includegraphics[width= 1\textwidth]{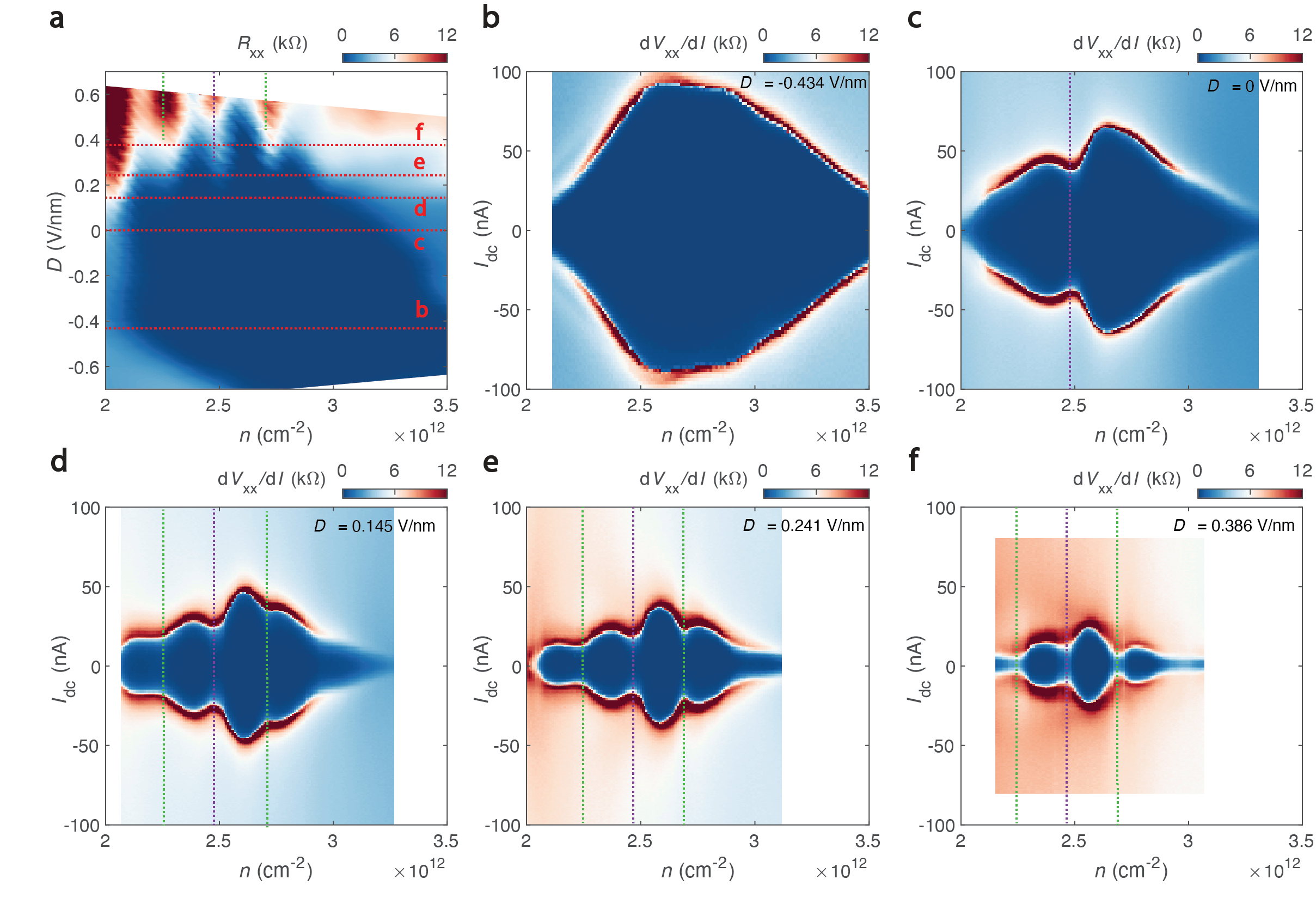}
  \caption{\textbf{Displacement field tunable superconductivity.} \textbf{a}, $n-D$ mapping of $R_{xx}$ at $T=240$~mK. \textbf{b - e}, $dV_{xx}/dI$ versus $I_{dc}$ and $n$ at different displacement fields. At a very large negative displacement field (\textbf{b}), superconductivity shows a single superconducting dome with the maximum critical current around 100~nA. As the displacement field increases, superconductivity weakens, and the maximum critical current decreases. More interestingly, the superconductor is gradually diced into small superconducting domes. At $D=0$~V/nm, the critical current shows a kink in the superconducting dome as marked by the purple dashed line where the carrier density is $n_{12}+n_{sm}$. This shows that the supermoiré lattice starts mediating the superconductivity. At higher displacement fields, there are more kinks of the critical current appearing as marked by green dashed lines in \textbf{d},  \textbf{e}, and \textbf{f}. The carrier density at these green lines corresponds to the half-filling of supermoiré mini-bands. This indicates the appearance of symmetry-broken phases in mini-flat bands. ($T=240$~mK)}
\label{Extended10}
\end{figure*}

\newpage
\clearpage

\section*{Supplementary materials}

\subsection*{Device homogeneity}

\begin{figure*}[h]
\renewcommand{\thefigure}{S1}
  \centering
  \includegraphics[width= 0.85\textwidth]{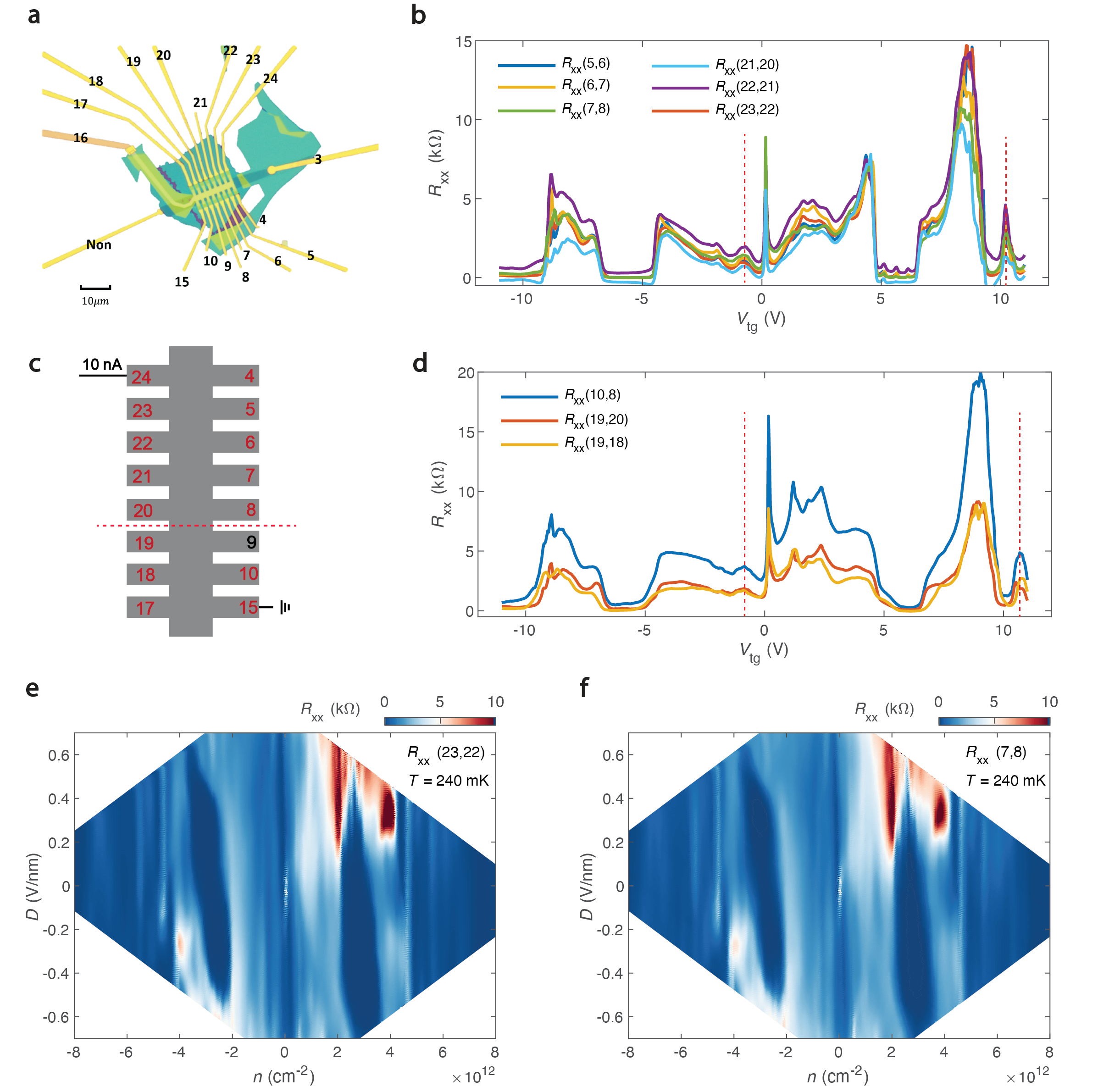}
  \caption{\textbf{Device pictures and homogeneity of the device.} \textbf{a}, optical microscope image of the device. \textbf{c}, the schematic of the measurement configuration. \textbf{b}, $R_{xx}$ as a function of $V_{tg}$ in the top region of the device shown in \textbf{c}. \textbf{d}, $R_{xx}$ as a function of $V_{tg}$ in the bottom region of the device shown in \textbf{c}. \textbf{d},\textbf{e}, $R_{xx}$ as a function of $n$ and $D$ by using contacts 23,22 and 7,8.} 
\label{sub1}
\end{figure*}

Fig.~\ref{sub1}a shows the optical microscope image of the device, and we label the contacts with numbers. The device has a metal top gate and a graphite bottom gate to independently tune the carrier density and displacement field. Fig.~\ref{sub1}b shows $R_{xx}$ as a function of $V_{tg}$ ($V_{bg}$ is kept at zero) of different contacts. The contact \textbf{24} acts as source contact where $I_s=10$~nA current is injected in, and contact \textbf{15} is the drain contact. We can see that the three pairs of contacts show almost the same results, the resistance peak near CNP and the peak at $n_{12}+n_{sm}$ as marked by red dashed line appear in the same position in these three regions. The results show the device is homogenous across this region (on the top part of the red dashed line in Fig.~\ref{sub1}c). Fig.~\ref{sub1}e,f are the $n-D$ mapping of $R_{xx}$ of another two pairs of contacts, which shows the same features as those in Fig.~\ref{fig1}b. We also notice that in a small region of the bottom part of the device, the twist angle of the top moiré pattern shifts slightly, as shown in Fig.~\ref{sub1}d. 

\subsection*{Checking alignment with hBN}

\begin{figure*}[h]
\renewcommand{\thefigure}{S2}
  \centering
  \includegraphics[width= 0.95\textwidth]{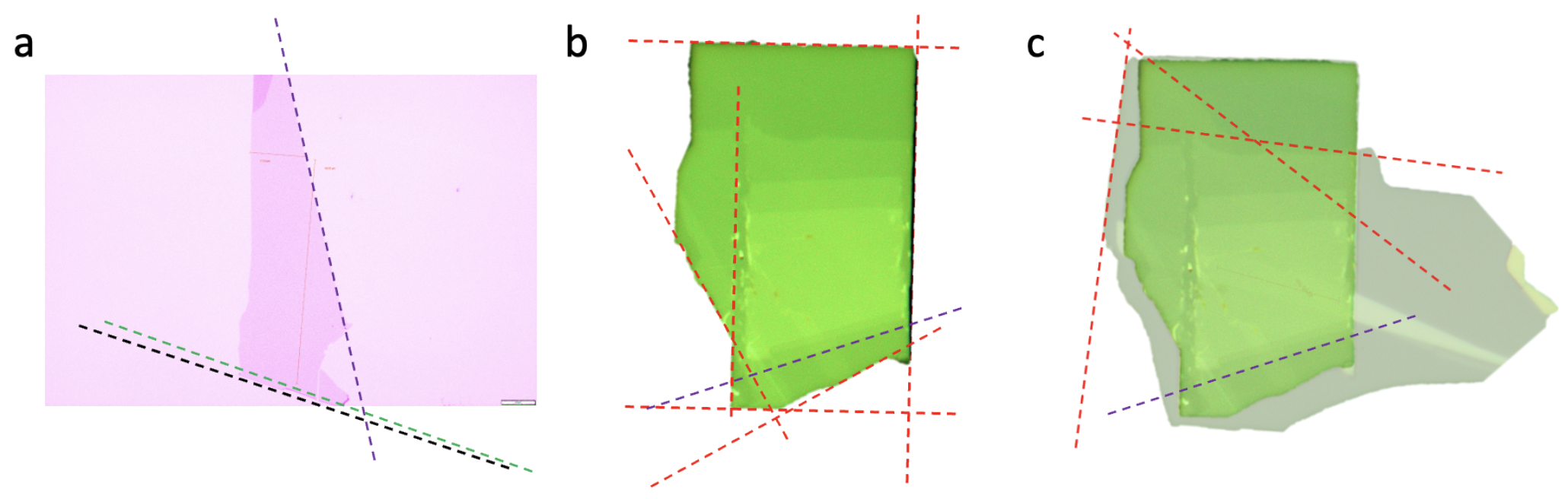}
  \caption{\textbf{Defining crystallographic axes of graphene and hBN.} \textbf{a,} The image of monolayer graphene we used for the twisted graphene. The purple line is the straight line we choose as the crystallographic axes of graphene. And the green/black lines are two lines we draw for the reference. The angle between purple line and green/black line is $58^\circ/59^\circ$. \textbf{b,} The red lines mark the crystallographic axes of top hBN and the purple line marks the crystallographic axes of graphene we choose. \textbf{c,} The red lines mark the crystallographic axes of bottom hBN and the purple line marks the crystallographic axes of graphene.}
\label{sub2}
\end{figure*}

\begin{figure*}[h]
\renewcommand{\thefigure}{S3}
  \centering
  \includegraphics[width= 0.95\textwidth]{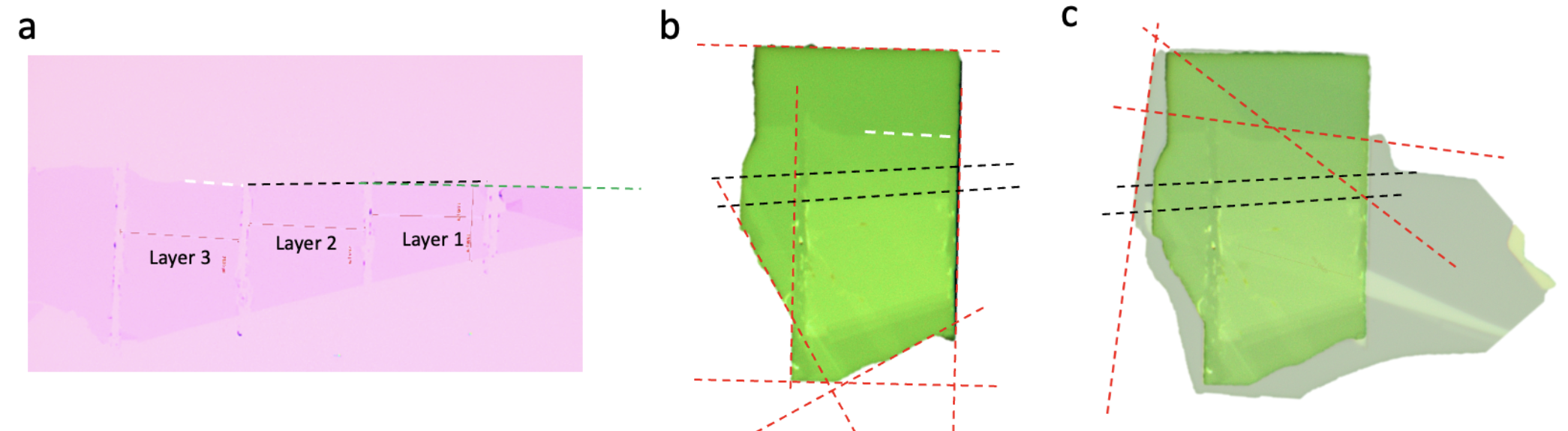}
  \caption{\textbf{Another edge of graphene.} \textbf{a,} The black marks the most straight part of the edge and the green dashed line marks the direction defined by the thick region of graphene. \textbf{b,} The red lines mark the crystallographic axes of top hBN and the two black dashed lines are the blacked lines defined in \textbf{a}. \textbf{c,} The red lines mark the crystallographic axes of bottom hBN and the two black dashed lines are the blacked lines defined in \textbf{a}.}
\label{sub3}
\end{figure*}

The alignment between hBN and graphene can create a moiré lattice due to the lattice mismatch. To rule out effects caused by such alignment, we examined the stacking configuration, as shown in Fig.~\ref{sub2} and Fig.~\ref{sub3}. Fig.~\ref{sub2}a shows the monolayer graphene used in our experiment. A very straight edge is highlighted by the purple lines. We introduced green and black dashed lines as guides for another graphene edge, and the angle between the purple and green/black lines is measured to be $58^\circ/59^\circ$. Another graphene edge lies between the green and black dashed lines. From this, we identified two graphene edges with a relative twist angle of $58^\circ \pm 0.5^\circ$, which is close to $60^\circ$. This allowed the use of the purple line to define the crystallographic axes of graphene. Fig.~\ref{sub2}b shows the top hBN, with its straight boundaries marked by red lines. The angles between these red lines are $30^\circ$, $60^\circ$, $90^\circ$, $120^\circ$, and $150^\circ$. This enables the use of the red dashed lines to represent the crystallographic axes of the top hBN. The purple line in Fig.~\ref{sub2}b corresponds to the crystallographic axes of graphene as defined in Fig.~\ref{sub2}a. The angle between the purple line shown in Fig.~\ref{sub2}b and the red dashed lines is measured to be $11^\circ$. This indicates that the twisted graphene is not aligned with the hBN, even when considering the measurement uncertainties in the angles. Similarly, in Fig.~\ref{sub2}c, the red lines represent the crystallographic axes of the bottom hBN, while the purple line denotes the graphene axes. The angle between them is $65^\circ$, further confirming that the twisted graphene is not aligned with the bottom hBN.

We also examined the situation using the other edge. Another edge is straight in the layer 1 and layer 2 region, as indicated by the black dashed line in Fig.~\ref{sub3}a. In Fig.~\ref{sub3}b, the red dashed line represents the crystallographic axes of the top hBN. The angle between the black dashed lines and the red dashed lines is $4^\circ/5^\circ$, confirming that the hBN is in the misalignment regime. Then if we choose the thick part as marked by the green line in Fig.~\ref{sub3}a to represent the crystallographic axes of graphene, the angle will be $3^\circ/4^\circ$. If we choose the other side to represent crystallographic axes of graphene, then the graphene has a chance to be aligned with hBN. But as we show in the above Fig.~\ref{sub2}, the edge we present in Fig.~\ref{sub2} more reliably represents the crystallographic axes of graphene. Fig.~\ref{sub3}c shows the relative orientation between another straight edge of graphene and the crystallographic axes of the bottom hBN. It is clear that the bottom hBN is not aligned with the another graphene edge.

Moreover, based on the results themselves, we can rule out the presence of a supermoiré lattice induced by hBN alignment. In our Landau fan diagram in Fig.~\ref{sub2} and the $n$–$D$ mapping shown in Fig.~\ref{sub6}, we observe two full fillings of the moiré lattice, denoted as $n_{12}$ and $n_{23}$. If the system were described by a conventional twisted trilayer graphene aligned with the top hBN, then $n_{12}$ would correspond to the full filling of the TTG moiré lattice, while $n_{23}$ would represent the full filling of the hBN moiré lattice. We can extract the moiré wavelength of hBN to be $7.89$ nm. Since the moiré effect from hBN diminishes rapidly as the moiré wavelength increases, $7.89$ nm suggests the system is in the misaligned regime, which won't alter band structure so much. Furthermore, we can extract twist angle of the TTG to be $1.328^\circ$, and the relative angle between TTG and hBN to be $1.52^\circ$, this combination won't give rise to a supermoiré length as $30$ nm.

\subsection*{Extracting capacitance between gates and graphene}

\begin{figure*}[h]
\renewcommand{\thefigure}{S4}
  \centering
  \includegraphics[width= 0.95\textwidth]{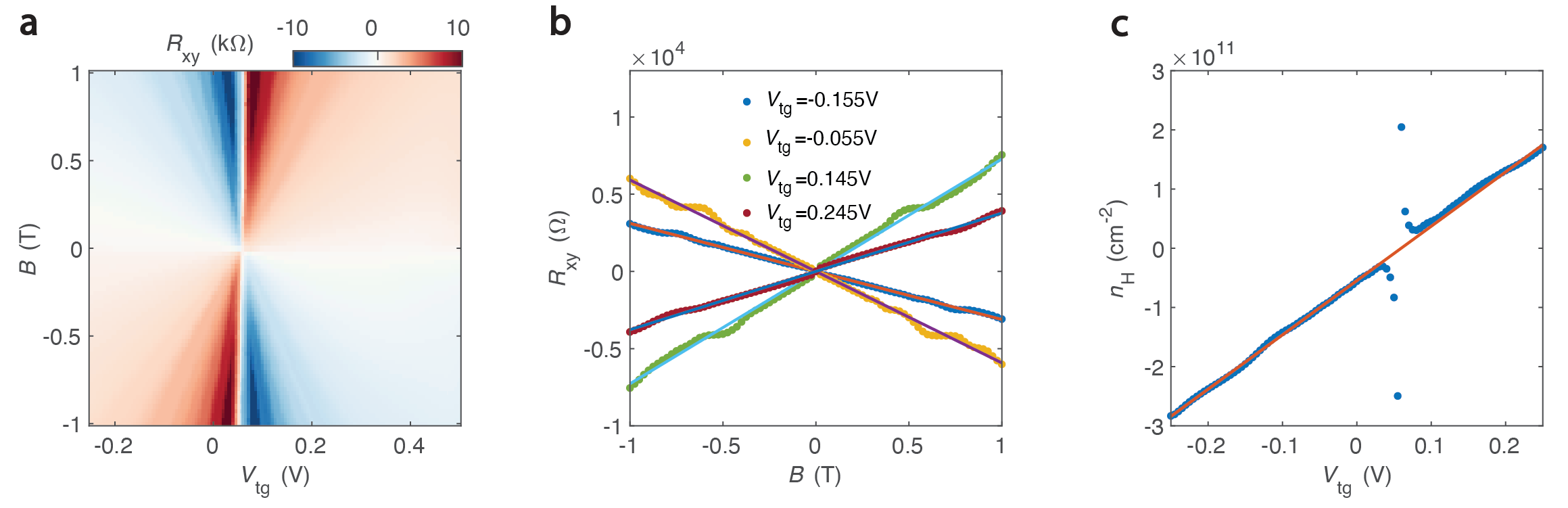}
  \caption{\textbf{Extracting capacitance between gates and graphene.} \textbf{a}, antisymmetric $R_{xy}$ as a function of $B$ and $V_{tg}$ near CNP when $D=0$~V/nm. \textbf{b}, line cuts of $R_{xy}$ in \textbf{a} and line fitting against $B$. \textbf{c}, the extracted Hall density as a functionof $V_{tg}$. ($T=240$~mK)}
\label{sub4}
\end{figure*}

The carrier density and the displacement field in the device can be expressed as: $n=(C_{tg}V_{tg}+C_{bg}V_{bg})/e$ and $D=(C_{tg}V_{tg}-C_{bg}V_{bg})/2\varepsilon_0$. In order to extract $n$ and $D$, the capacitances between gates and graphene are needed. Near CNP, Hall density is equal to the carrier density in the device, and the Hall density can be extracted from Hall resistance by the relationship $R_{xy}=B/ne$. Fig.~\ref{sub4}a shows the antisymmetric $R_{xy}=(R_{xy}(B)-R_{xy}(-B))/2$ as a function of $V_{tg}$ when $D=0$~V/nm. Fig.~\ref{sub4}b shows several line cuts from Fig.~\ref{sub4}a, and the line fit of $R_{xy}$ against $B$ will give us the Hall density. Fig.~\ref{sub4}c shows the Hall density versus $V_{tg}$. Because we fix the displacement field at zero, so we have a relationship: $C_{tg}V_{tg}=C_{bg}V_{bg}$, and the carrier density can be expressed as $n=2C_{tg}V_{tg}/e$. By performing a linear fit of $n_H$ as a function of $V_{tg}$, we can extract the capacitance per unit area between the top gate and graphene, $C_{tg}$. From the features appearing at the charge neutrality point in the $R_{xx}(V_{tg}, V_{bg})$ map, we obtain the relationship between $C_{tg}$ and the bottom gate capacitance $C_{bg}$, allowing us to determine $C_{bg}$.

\subsection*{Determine full filling of two moiré lattices}

\begin{figure*}[h]
\renewcommand{\thefigure}{S5}
  \centering
  \includegraphics[width= 0.9\textwidth]{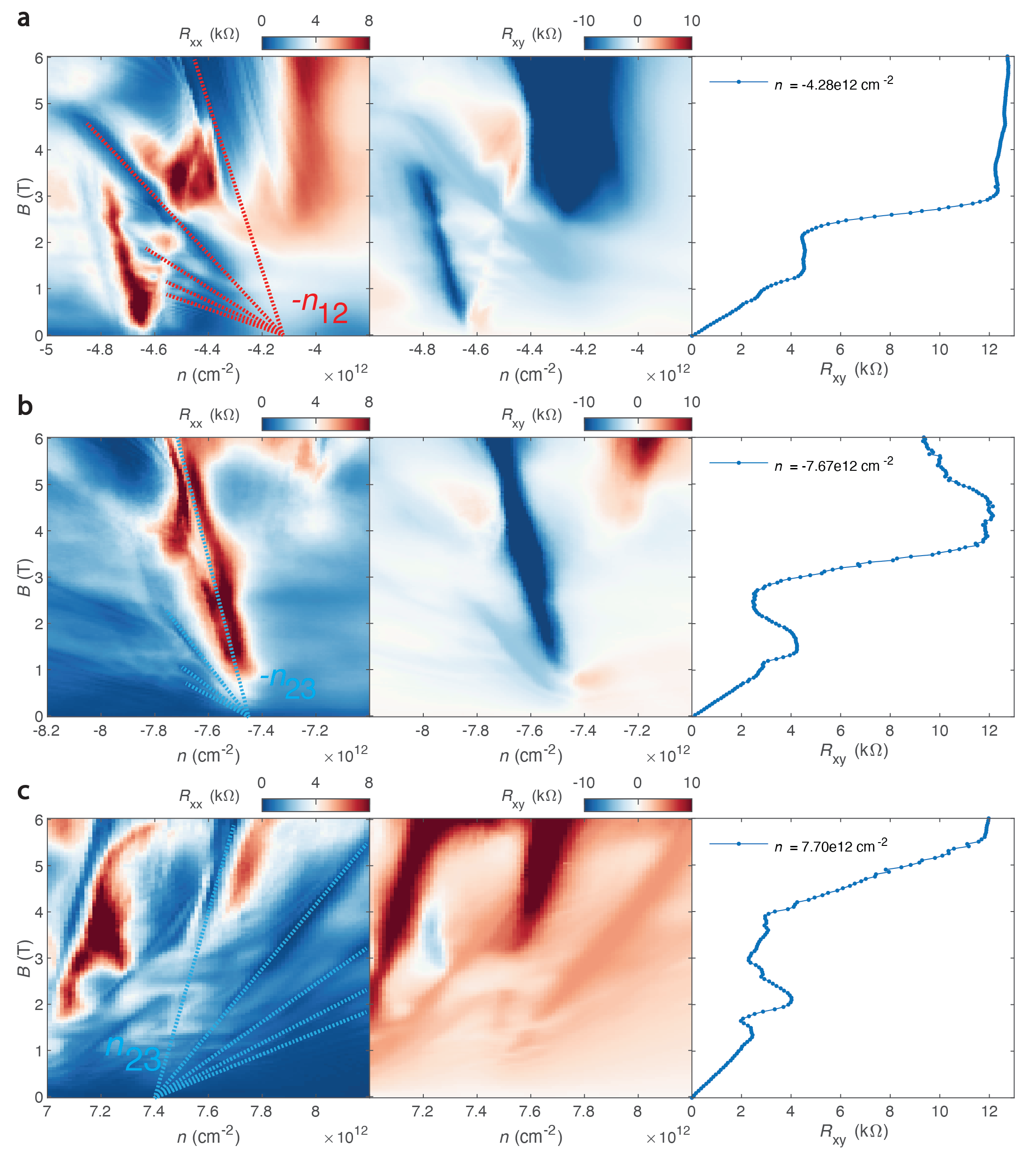}
  \caption{\textbf{Landau fan diagrams near full fulling of two moiré lattice.} \textbf{a}, Landau fan diagram of $R_{xx}$ and $R_{xy}$ at $D=0$~V/nm on the hole-doped side near the full filling of the top moiré lattice. The red lines mark the Landau levels originating from $-n_{12}$ and show a sequence of $-2, -6, -10\ldots$. \textbf{b} (\textbf{c}), Landau fan diagram of $R_{xx}$ and $R_{xy}$ at $D=0$~V/nm on the hole-doped (electron-doped) side near the full filling of the bottom moiré lattice. The blue dashed lines show Landau levels stemming from $-n_{23}$ and $n_{23}$ and has a equency of $-2, -6, -10\ldots$ ($2, 6, 10\ldots$). ($T=240$~mK)}
\label{sub5}
\end{figure*}

In Fig.~\ref{fig1}d,e, many sets of Landau levels are observed, and we need to identify those originating from the full filling of the top and bottom moiré lattice to determine the twist angles. In the mirror symmetry TTG device, only the Dirac band will contribute to the electronic properties when the fermi level is tuned into the gap between the flat band and remote conduction (valence) bands. So, Landau levels can stem from the Dirac band, and all the Landau levels will extend to the full filling of the flat band in zero magnetic fields. In Fig.~\ref{fig1}d,e, we can find four regions showing clear Landau levels. Zoom-in of one region is shown in Fig.~\ref{fig3}a and Fig.~\ref{Extended5}a. Fig.~\ref{sub5} shows a zoom-in of the Landau fan diagram in the other three regions. In Fig.~\ref{sub5}a, robust Landau levels are observed and marked by the red dashed lines. The Landau levels have a sequence of $-2, -4, -6\ldots$, so we identify the full filling of the top moiré lattice on the hole-doping side $-n_{12}$. Using a similar way, we can also find the full filling of the bottom moiré lattice on both the hole-doped side and electron-doped side as shown in Fig.~\ref{sub5}b,c. 

Fig.~\ref{sub6} show the $n-D$ mapping of $R_{xx}$ and $R_{xy}$ at $B=1$~T. We marked the full filling of the bottom moiré lattice and top moiré lattice with red and blue dashed lines. We can see that the full filling state is more robust at the positive (negative) displacement field side on the electron-doped (hope-doped) side. This is due to the positive (negative) displacement field polarizing electrons (holes) on the top of the device.

\begin{figure*}[h]
\renewcommand{\thefigure}{S6}
  \centering
  \includegraphics[width= 0.9\textwidth]{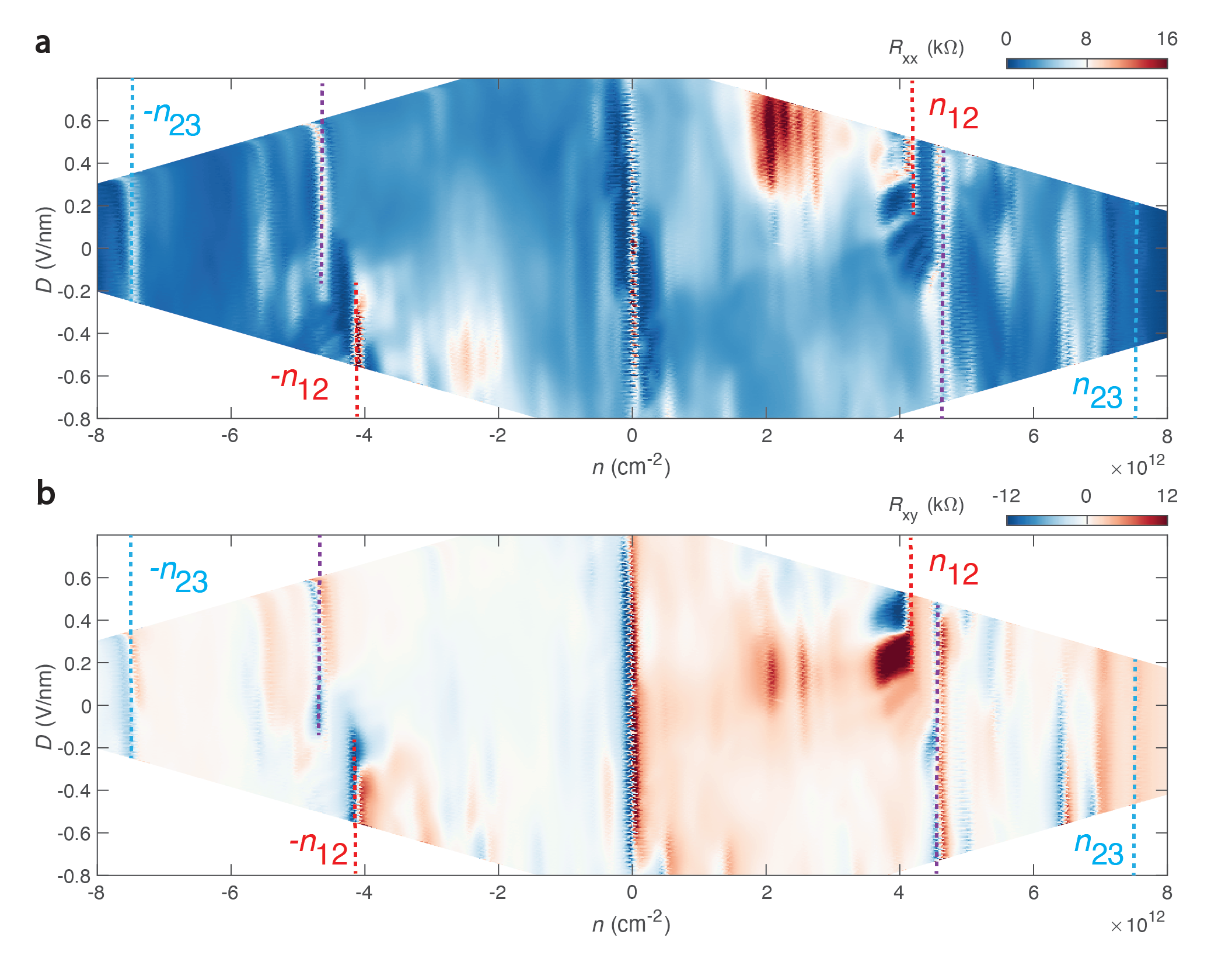}
  \caption{\textbf{$n-D$ mapping of $R_{xx}$ and $R_{xy}$ when B=1T.} The blue and red lines denote the full filling of the bottom moiré lattice and top moiré lattice. And the black lines mark the state at $n_{12}+n_{sm}$.}
\label{sub6}
\end{figure*}

\begin{figure*}[h]
\renewcommand{\thefigure}{S7}
  \centering
  \includegraphics[width= 0.8\textwidth]{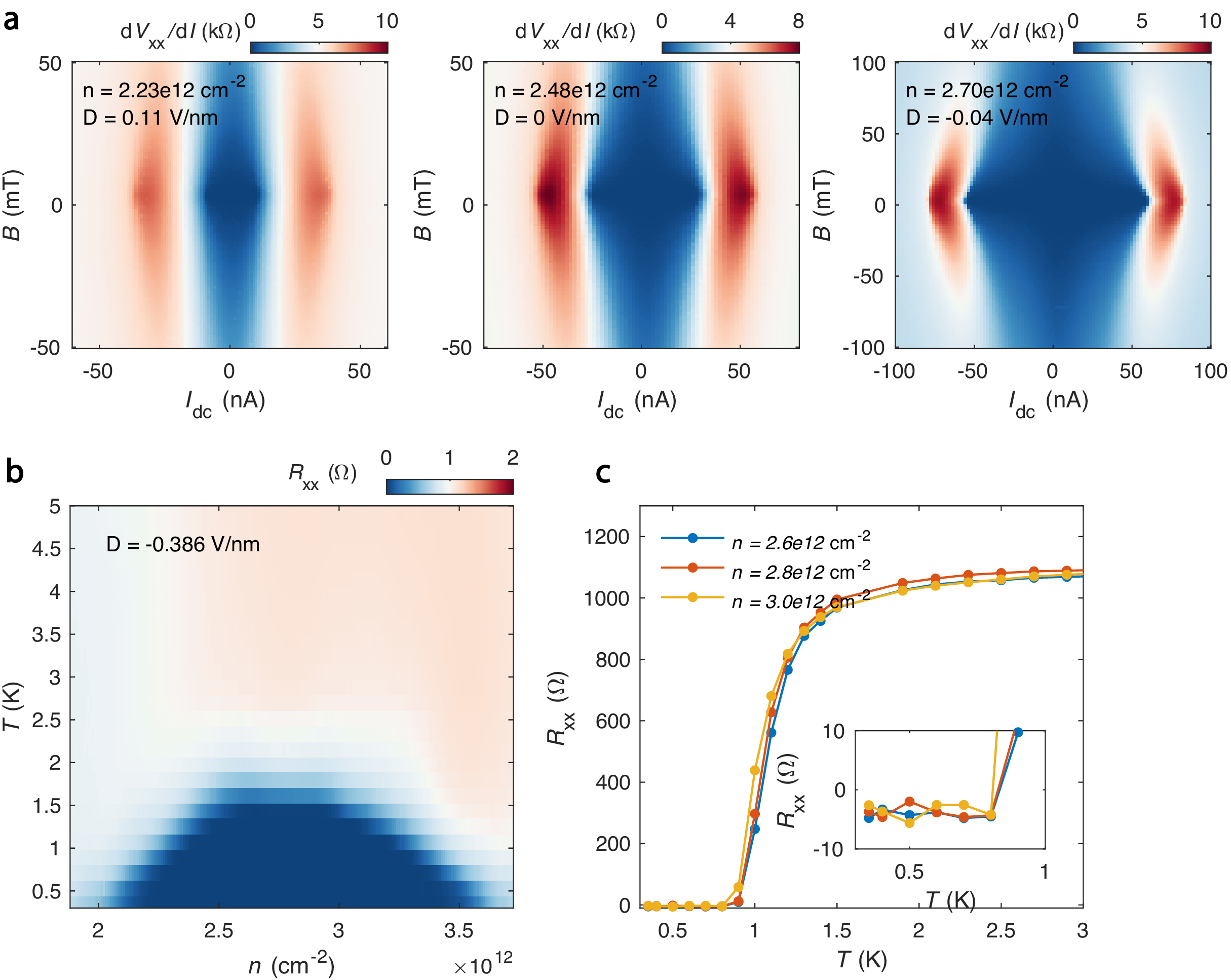}
  \caption{\textbf{Magnetic field dependence and temperature dependence of superconductivity.} \textbf{a,} $dV_{xx}/dI$ versus magnetic field and $I_{dc}$ at three different points in the $n-D$ mapping. \textbf{b,} Temperature dependence of $R_{xx}$ at $D=-0.386$ V/nm. \textbf{c,}Line cuts of $R_{xx}$ versus $T$ and the zoom in shows the resistance below critical temperature.}
\label{sub6}
\end{figure*}

\end{document}